\documentclass[11pt]{article} 

\usepackage{amsmath,amsthm,latexsym,amssymb,amsfonts,epsfig}

\newcommand{\be}{\begin{equation}}
\newcommand{\ee}{\end{equation}}
\newcommand{\ba}{\begin{eqnarray}}
\newcommand{\ea}{\end{eqnarray}}

\title{{\sf Hamiltonian renormalisation VI:}\\
 {\sf Parametrised field theory on the cylinder}} 
\author{
{\sf T. Thiemann}$^1$\thanks{{\sf 
thomas.thiemann@gravity.fau.de}},
{\sf E.-A. Zwicknagel}$^1$\thanks{{\sf 
ernst-albrecht.zwicknagel@gravity.fau.de}}\\
\\
{\sf $^1$ Inst. for Quantum Gravity, FAU Erlangen -- N\"urnberg,}\\
{\sf Staudtstr. 7, 91058 Erlangen, Germany}\\
}
\date{{\small\sf \today}}

\makeatletter
\@addtoreset{equation}{section}
\makeatother

\begin{document} 

\maketitle

{\sf

\begin{abstract}
Hamiltonian Renormalisation, as defined within this series of works, 
was derived 
from covariant Wilson renormalisation via Osterwalder-Schrader reconstruction.
As such it directly applies to QFT with a true (physical) Hamiltonian bounded 
from below. The validity of the scheme was positively tested for free QFT 
in any dimension with or without Abelian gauge symmetries of Yang-Mills type.

The aim of this Hamiltonian renormalisation scheme is to remove quantisation
ambiguities of Hamiltonians in interacting QFT that remain even after UV and 
IR regulators are removed as it happens in highly non-linear QFT such as 
quantum gravity. Also, while not derived for that case,
the renormalisation flow formulae can without change 
also be applied to QFT without a single true Hamiltonian but rather an 
infinite number of Hamiltonian constraints. In that case a number of 
interesting questions arise: 1. Does the flow reach the correct fixed 
point also for an infinite number of ``Hamiltonians'' simultaneously?
2. As the constraints are labelled by test functions, which in presence of 
a regulator are typically regularised (discretised and of compact support),
how do those test functions react to the flow? 3. Does the 
quantum constraint algebra, which in presence of a regulator is expected 
to be anomalous, close at the fixed point? 

These questions should ultimately be addressed in quantum gravity. Before 
one considers this interacting, constrained QFT it is well
motivated to consider a free, constrained QFT where the fixed point is 
explicitly known. In this paper we therefore address the case of 
parametrised field theory for which the quantum constraint algebra
coincides simultaneously with the hypersurface deformation algebra of 
quantum gravity (or any other generally covariant theory) and the Virasoro
algebra of free, closed, bosonic 
string theory or other CFT to which the results of this paper apply 
verbatim.

The central result of our investigation is that finite resolution 
(discretised) constraint algebras {\it must not close} and that anomaly 
freeness of the continuum algebra is encoded in the 
convergence behaviour of the renormalisation flow.
\end{abstract}

\section{Introduction}  
\label{s1}

Interacting QFT typically have to be constructed: One first defines a 
regulated theory (with both UV and IR regulator present) and then tries 
to remove the regulator thereby renormalising the bare parameters 
(i.e. redefining them in terms of measured parameters and regulators).
That procedure of {\it constructive QFT}, even if successful in the sense 
that the unregulated, non-perturbative theory is well defined, may yet be 
ambiguous, 
i.e. it may keep a memory of which regularisation procedure was applied.
We will refer to such ambiguities as {\it quantisation ambiguities}. 
One expects this problem the more likely to occur the more non-linear
the theory is. An extreme case is quantum gravity whose Einstein-Hilbert 
action depends non-poynomially on the metric field. 

Such ambiguities are not severe if they can be encoded by a finite number 
of (so called relevant) parameters. They could be fixed by a finite number
of experiments and thus lead to a predictive theory. However, if that 
parameter space is infinite dimensional, the theory is not predictive. 
To make it predictive, the number of free parameters must be downsized 
to a finite dimensional manifold. To achieve this, one imposes a restriction
on the family of regulated theories: they must qualify, at finite regulator,
as the coarse grained versions of a continuum theory at a resolution 
defined by that regulator. For instance a Euclidian QFT maybe defined by a 
family of measures $\mu_r$ where $r$ denotes the regulator.  The measure 
$\mu_r$ knows how to integrate functionals of the Euclidian quantum field 
smeared with test functions that are restricted up to resolution $r$.
Thus, in order to produce unambiguous results, for any finer resolution 
$r'<r$ we must have $[\mu_{r'}]_r=\mu_r$, i.e. since the quantum field 
tested at resolution $r$ can be written in terms of the quantum field at 
resolution $r'$ we can use $\mu_{r'}$ instead of $\mu_r$ to integrate
functions restricted to resolution $r$.      

This is so called {\it cylindrical consistency} 
is basically the idea of Wilson renormalisation \cite{1}. A cylindrically 
consistent family of measures $\mu_r$
in turn defines a continuum measure $\mu$ that can integrate the quantum field
at any resolution under 
rather mild assumptions \cite{2}. From a practical viewpoint, the cylindrical
family is then sufficient because in reality one never considers physical 
processes at infinite resolution, thus the explicit construction of $\mu$
is not needed. Now in constructive QFT one typically starts with an 
initial family $\{\mu^{(0)}_r\}_{r\in R}$ where $R$ is the regulator manifold.
It typically comes with an in principle infinite number of parameters $p\in P$
that 
enter via the discretisation freedom 
of the classical theory (action) that one starts 
from (e.g. next neighbour, next to next neighbour ,... terms in the 
Laplacian). Even if the limit $\mu^{(0)}:=\lim_{r\to 0}\; \mu^{(0)}_r$ exists
as a measure, it will typically retain a non-trivial dependence on {\it all} 
``directions'' of the parameter manifold $P$. Therefore it is natural 
to {\it improve} the initial family and define a sequence of families
$n\in \mathbb{N}_0\;\mapsto \; \{\mu{(n)}_r\}_{r\in R}$ by 
$\mu^{(n+1)}_r:=[\mu^{(n)}_{\kappa(r)}]_r$ where $\kappa(r)<r$ maps to 
a finer resolution. This defines a flow of measure families which may 
have a fixed point $\{\mu^\ast_r\}_{r\in R}$ which by construction 
is consistent at least with respect to the {\it coarse grainings} 
$\kappa(r)\to r$. Experience shows that this usually also makes the 
fixed point family consistent with respect to all pairs $r'<r$. In
the course of this process, it may happen that all but a finite (relevant)
directions in $P$ have been fixed to a fixed value. In that case we 
say that the QFT has been non-perturbatively renormalised to a predictive   
QFT.

These ideas were first formulated in quantum statitistical field theory 
(i.e. Euclidian field theory \cite{3}) using path integral methods.
Using Osterwalder-Schrader (OS)
reconstruction one can also translate them into the Hamiltonian language 
\cite{4} (see also \cite{5} for closely related earlier Hamiltonian 
renormalisation schemes and references therein). The validity of \cite{4}
has been tested in free field theories without \cite{6} and with \cite{7}
Abelian gauge symmetry of Yang-Mills type. The motivation for \cite{4}
is actually its application in Hamiltonian quantum gravity, specifically
in its Loop Quantum Gravity (LQG) incarnation \cite{8}. Since the classical
Einstein-Hilbert action is non-polynomial in the metric field, the 
quantisation ambiguity problem is expected to be especially severe in this 
case. Indeed, quantum gravity is not perturbatively renormalisable which 
motivates the non-perturbative path integral renormalisation programme known 
as asymptotic safety \cite{9}. In the Hamiltonian setting, while 
it is possible to rigorously define the Hamiltonian constraint operators
\cite{10} they suffer from quantisation ambiguities so that a Hamiltonian
renormalisation thereof is well motivated \cite{11,7a}.        
      
At first it may look strange why an OS motivated Hamiltonian 
renormalisation scheme should apply at all to quantum gravity: OS 
reconstruction delivers a Hamiltonian operator $H$ bounded from below on 
a Hilbert space $\cal H$ and a ground state $\Omega\in {\cal H}$. However,
canonical quantum gravity does not come with a Hamiltonian but rather 
an infinite number of Hamiltonian constraints $C(N)$ on a 
Hilbert space ${\cal H}'$ where $N$ are test 
functions (called lapse functions). For no choice of $N$ are these bounded 
from below and rather than the spectrum of $H$ on $\cal H$ 
one is interested in the joint kernel of the $C(N)$ on ${\cal H}'$ 
defining the physical
Hilbert space $\cal H$ which does not coincide with 
${\cal H}$ and is typically not a subspace thereof 
(typically it is a space of distributions on a dense subspace of ${\cal H}'$).
However, on the one hand 
it is possible to cast quantum gravity into the framework 
of an ordinary quantum Hamiltonian system by using Hamiltonian constraint gauge
fixings \cite{12}. In this reduced phase space approach one then retains 
a physical Hamiltonian directly on the physical Hilbert space $\cal H$.

On the other hand, it turns out that the Hamiltonian renormalisation flow,
while derived from the OS renormalisation scheme, can formally be applied 
also to more than one operator and in particular also those 
which are not bounded from below, certainly but not necessarily 
when they share a common ground state
$\Omega$. This observation allows the attractive perspective to monitor the 
fate of the commutator algebra of the $C(N)$ during the renormalisation 
process which is not possible in the 
reduced phase approach where the $C(N)$ are solved classically. Classically
we have the closed hypersurface deformation algebra \cite{13} 
$\{C(M),C(N)\}=C(f(M,N))$ where $f(M,N)$ are new test functions which 
in more than two spacetime dimensions or with density weight different from 
two also depend on the metric. This
fact makes it especially difficult to turn this into an {\it anomaly free}
constraint operator commutator $[C(M),C(N)]=i\hbar$ ``$C(f(M,N))$'' because 
of the ordering problem involved in $C(f(M,N))$ \cite{14}. Indeed, the 
development of \cite{10} can be interpreted as saying that 
$[C(M),C(N)]=i\hbar C(\tilde{f}(M,N))$ closes with the correct ordering
(i.e. the kernel of the $C(f(M,N))$ is contained in that of the $C(N)$)
but with the wrong ``structure functions'', that is, the operators 
$\tilde{f}(M,N)$ do not qualify as the quantisation of $f(M,N)$.      
To improve on this state of affairs, one may modify the quantisation 
of the $C(N)$ without resorting to renormalisation methods, an ambitious
very interesting  
programme that is now in motion \cite{15} and to which the developments of 
the current paper maybe viewed as complimentary, see especially the 
parametrised field theory application of that programme \cite{16}
(and also \cite{17} where qualitatively similar results were obtained 
without changing the notion of convergence of regulated operators as defined
in \cite{10}). 

More in detail, the Hamiltonian renormalisation flow works with a family
of triples $({\cal H}_r,\; H_r, \;\Omega_r)$ where ${\cal H}_r$ is a 
Hilbert space, $H_r$ a self-adoint operator on ${\cal H}_r$
(bounded from below if coming from an OS measure) and $\Omega_r$ is a 
ground state of $H_r$ i.e. $H_r\;\Omega_r=0$. The regulator labels $r$ belong
to partially ordered and directed set $R$. Given isometric embeddings 
$J_{r r'}:\; {\cal H}_r\to {\cal H}_{r'};\;r<r'$
to be constructed subject to the consistency condition 
$J_{r' r^{\prime\prime}} \; J_{r r'}=J_{r r^{\prime\prime}};\;\;
r<r'<r^{\prime\prime}$ and that ensure $J_{r r'}\;\Omega_r =\Omega_{r'}$
one defines the 
inductive limit Hilbert space $\cal H$ by a standard construction \cite{18}.
Moreover, at the fixed point, the $H_r$ form a consistently defined 
family of quadratic 
forms $H_r=J_{r r'}^\dagger\; H_{r'}\; J_{r r'},\;\;r<r'$ defining a 
continuum form $H$. That form may or may not define a self-adjoint operator
on $\cal H$ and in particular is in general not to be confused with the 
inductive limit of the $H_r$ which is not granted to exist.  \\
\\
In extending this framework to more than one (in field theory, even 
an infinite number of) operators, we face 
several new questions:\\
1.\\ 
We start with an initial family 
$C_r^{(0)}(N)$ of operators on an initial family of Hilbert spaces 
${\cal H}^{(0)}_r$, one for each resolution scale $r$ and one for each 
continuum smearing function $N$. The origin of $r$ typically comes 
from a discretisation of the continuum field $\phi$ and conjugate momentum 
$\pi$ in terms of coarse grained variables $\phi_r,\pi_r$ and substituting
them for $\phi,\pi$ in the expression for $C(N)$. Does this automatically
induce a discretisation $N^{(0)}_r$ of $N$ as well? If not, should one supply
one by hand or leave $N$ in its continuum form?\\
2.\\
Is it possible or necessary to find a common zero eigenvector 
$\Omega^{(0)}_r\in {\cal H}^{(0)}_r$ of the
$C^{(0)}_r(N)$ or     
$C^{(0)}_r(N_r^{(0)})$ independent of $N$ or $N^{(0)}_r$? This is far 
from  trivial: while the classical continuum constraints form a closed
Poisson algebra of real functions, there is no reason to 
take it for granted that the algebra of the 
discretised 
$C^{(0)}_r(N)$ or $C^{(0)}_r(N^{(0)}_r))$ closes under taking commutators.
In fact this is most likely not the case because typically the classical 
constraint algebra rests on the validity of the Leibniz rule for 
partial derivatives. However, discretised derivatives do not obey 
the Leibniz rule \cite{19}. Thus not only can these constraints 
not be simultaneously diagonalised, it may even be that their joint kernel 
just consists of the zero vector. In that case, we have to assume that 
there exists at least a cyclic vector $\Omega^{(0)}_r$ for the algebra 
of operators under consideration in the common dense 
domain of all constraints.\\     
3.\\      
Given that $\Omega^{(0)}_r$ can be found, one can proceed as in the case of 
just one Hamiltonian operator and construct a sequence of families of 
Hilbert spaces ${\cal H}^{(n)}_r$ and isometric 
injections $J^{(n)}_{r r'}:\; {\cal H}^{(n+1)}_r\to {\cal H}^{(n)}_{r'}$
for $r'<r$ such that $J^{(n)}_{r r'} \Omega^{(n+1)}_r=\Omega^{(n)}_{r'}$.
The isometry requirement translates into flow equations for the Hilbert 
space measures $\nu^{(n)}_r$ underlying 
${\cal H}^{(n)}_r=L_2(Q_r,d\nu^{(n)}_r)$ where
$Q_r$ is a flow invariant model configuration space. Assuming that 
a fixed point $J_{r r'}$ of this flow of isometric injections can be found
(equivalently, a cylindrically consistent measure family $\nu_r$) one can
construct a continuum Hilbert space $\cal H$ as the inductive limit of the 
${\cal H}_r=L_2(Q_r,d\nu_r)$. In tandem, one constructs a flow of families 
quadratic forms 
$C^{(n+1)}_r(N):=[J^{(n)}_{rr'}]^\dagger C^{(n)}_{r'}(N) 
J^{(n)}_{r r'}$ one for each $N$. Can one arrange that all of them 
flow into a fixed point whatever choice of $N$ is made? Or should one 
rather also let the discretised smearing functions flow according to 
$C^{(n+1)}_r(N^{(n+1)}_r):=[J^{(n)}_{rr'}]^\dagger C^{(n)}_{r'}(N^{(n})_{r'}) 
J^{(n)}_{r r'}$?\\
4.\\
Suppose that a simultaneous fixed point family $C_r(N)$ or $C_r(N_r)$ 
can be found. Then by construction 
$C_r(N)=J_r^\dagger C(N) J_r$ or $C_r(N_r)=J_r^\dagger C(N) J_r$
where $J_r:\; {\cal H}_r\to {\cal H}$ is the isometric embedding granted to 
exist by the inductive limit construction. Is it true that $C(N)$ 
is no longer plagued by an infinite number of quantisation 
ambiguities? Is it true that the algebra of commutators of $C(N)$ is 
non-anomalous? Note that it is not clear that the commutators can even be 
computed because $C(N)$ is just a quadratic form. \\
5.\\
Assuming that these questions can be answered in the affirmative, how does 
one recognise anomaly freeness at finite resolution? Note that the $C_r(N)$
will most certainly not close under forming commutators even if the 
$C(N)$ do because \cite{11}  
\be \label{1.1}
[C_r(M),C_r(N)]=
J_r^\dagger [C(M) P_r C(N)-C(N) P_r C(M)] J_r
\ee
where 
$P_r:=J_r J_r^\dagger$ is a projection in ${\cal H}$. 
{\it It is therefore generically
not expected that the finite resolution projections of the constraints form 
a closed algebra}! However, given closure in the continuum, 
we may rewrite (\ref{1.1}) as
\be \label{1.2}
[C_r(M),C_r(N)]=i\hbar\;C_r(f(M,N))
-J_r^\dagger [C(M)\; (1_{{\cal H}}-P_r)\; C(N)
-C(N)\; (1_{{\cal H}}-P_r]\; C(M)) J_r
\ee
and the anomalous term naively vanishes as $r$ is removed and $P_r$ becomes 
$1_{{\cal H}}$. This, when supplied by a suitable operator 
topology of convergence, 
may serve as a pratical guide towards proving anomaly 
freeness even if one cannot determine the continuum operator $C(N)$ in closed 
form.\\
\\
It would be very interesting to find necessary and sufficient conditions 
under wich the above questions can be answered in the affirmative. In this 
paper we confine ourselves to the much easier task to illustrate and 
work out the 
catalog of questions and answers for the case of parametrised massless 
Klein-Gordon field theory 
in 1+1 spacetime dimensions.\\
\\
The architecture of this paper is as follows:\\
\\
In section \ref{s2} we briefly review 1+1 dimensional PFT following the 
notation of \cite{17}. We treat both the classical and quantum theory.

In section \ref{s3} we specialise the general framework of \cite{4,11} 
to PFT. We choose as regulator space a nested system of square lattices.
Here we learn the first important lesson from the present work: 
The constraint operators are ill defined on the dense domain of finite 
resolution subspaces generated by the discretised Weyl algebra unless 
the test functions that enter that Weyl algebra and which define the 
renormalisation flow display at least a minimal amount of smoothness.
This issue did not arise in the works \cite{6} because there the 
renormalisation could be phrased in terms of the covariance of the 
Gaussian measure which decays sufficiently fast at infinity in momentum 
space even when smeared against the discontinuous test functions used. 
However in PFT we also need inverse powers of that covariance. This 
obervation triggered the work \cite{27} where we generalise 
\cite{11} in a natural way to a generalised Multi-Resolution Analysis (MRA) 
based renormalisation flows of which there are even smooth candidates 
thus removing the afore mentioned obstacle. In fact, \cite{6} turns 
out to be a special case of \cite{27} as \cite{6} is based on the 
so-called Haar MRA. On the other hand, as 
the convergence to the continuum via sequences of discontinuous or smooth 
functions should not affect the continuum fixed point theory, we also 
offer an equivalent solution to the just mentioned smoothness problem
within the Haar MRA class based on zeta function regularisation which is 
a common tool in conformal field theories (CFT) such as PFT.     

In section \ref{s4} we show that there exists a well motivated discretisation
of the PFT constraints. Clearly, due to the central term in the 
Virasoro algebra there does not exist a single vector in the joint point 
kernel of all constraints, not even in the continuum. However, there does 
exist a preferred cyclic vector in the common dense domain of all constraints 
which serves as a substitute, both in the continuum and at finite 
resolution. We can then proceed similar as in \cite{6,7} and 
compute the flow and fixed point of the corresponding Hilbert space measures.

In section \ref{s5} we compute the flow of the constraint operators.
We show that the first option of 
leaving the smearing functions $N$ untouched (not discretised 
by hand) does not induce a canonical discretisation of the smearing 
functions of the constraints.  
On the other hand, using the coarse graining map that is used to compute 
the flow of measures, vacua and constraints to discretise their smearing 
functions by hand does lead to a cylindrically 
consistent system (under change of resolution) of constraints.  

In sections \ref{s6} and \ref{s7} 
we compute the algebra of constraints at finite 
resolution and illustrate the behaviour (\ref{1.1}), (\ref{1.2}).
It is at this point that we learn the second most important lesson from 
the present work
when trying to show that the discrete 
algebra converges to the continuum algebra in the weak operator 
topology:\\
i. When working with non-discretised constraint smearing functions, there 
is just one correction to the continuum algebra at finite resolution 
indicated in (\ref{1.1}) and (\ref{1.2}). However, when additionally 
discretising the constraint smearing function by hand, an additional 
correction arises.\\
ii. Convergence to zero of the first correction requires a minimal 
amount of smoothnees of the test functions of the Weyl algebra for reasons 
similar as mentioned before concerning the domain of definition of 
the constraints. \\
iii. Convergence to zero of the second correction requires sufficient 
smoothness of the discretised smearing function $N$ of the constraints,
which is of course not surprising because the Virasoro algebra depends on 
third order (Schwartzian)
derivatives of those smearing functions. \\
We establish convergence using for instance the Dirichlet flow of 
\cite{27} rather than the Haar flow of \cite{6}. 

In section \ref{s8} we summarise and conclude our findings for this model
which presents the next logical step in the research programme 
started in \cite{4,6,7,11}.\\
\\ 
The most important lessons learnt from the present 
work are as follows:\\
A. Finite resolution constraints {\it typically do not close}.\\
B. This is {\it no problem at all, in fact it would be physically wrong:} 
It just displays the mathematical fact 
that the constraints typically are not block diagonal w.r.t. different 
resolution Hilbert subspaces. The failure to close is {\it no anomaly 
but a finite resolution artefact.}\\
C. Whether the {\it continuum algebra} closes, i.e. is free of 
anomalies {\it can be checked using finite resolution analysis}: The 
finite resolution artefact should converge to zero. This is of practical 
importance because in more complicated theories one will hopefully be able to 
construct the theory at finite resolution but perhaps computing the 
infinite resolution (continuum) theory may be too hard but also unnecessary 
as measurements always have finite resolution.

\section{Brief review of PFT}
\label{s2}

This section mainly serves to introduce our notation and follows \cite{17}.
See \cite{17} for more information and references therein. See also 
\cite{15a} for more details on the quantisation of PFT using classically
equivalent constraints for which the quantum anomaly is formally a co-boundary
so that it can be (formally - i.e. modulo showing existence of corresponding
Hilbert space representations) 
absorbed into a non-central quantum correction of 
the constraints. See \cite{15b} for renormalisation of closely related 
(fermionic) CFT's. 

\subsection{Classical Theory}
\label{s2.1}

The spacetime is the infinite cylinder $Z_R=\mathbb{R}\times C_R$ 
where $C_R$ is the circle of radius
$R$ with Minkowski metric $\eta$=diag$(-1,1)$ and Cartesian coordinates 
$T:=X^0\in \mathbb{R},\; X:=X^1\in [0,2\pi R)$. We introduce another 
cylinder $Z$ of unit radius $Z=\mathbb{R}\times S^1$ with coordinates 
$(x^0=t,x^1=x)$ and consider the diffeomorphism 
$\varphi:\;Z\to Z_R;\; (t,x)\mapsto 
(T(t,x),X(t,x)$ upon which $T,X$ become fields on $Z$. Note that $T$ is 
periodic $T(t,x+1)=T(t,x)$ while $X$ is an angular variable 
$X(t,x+1)=X(t,x)+2\pi\;R$. 

The action of the massless Klein-Gordon field $\phi$ on $Z_R$
\be \label{2.1}
S=-\frac{1}{2} \; \int_{Z_R}\; d^2X\; \eta^{AB}\; \phi_{,A}\; \phi_{,B}
\ee 
is pulled back by above diffeomorphism and yields via 
$\phi=\varphi^\ast \; \Phi$ the PFT action
\be \label{2.2}
S=-\frac{1}{2} \; \int_{Z}\; d^2x\; |\det(g)|^{1/2}\;
g^{\alpha\beta}\; \Phi_{,\alpha}\; \Phi_{,\beta};\;\;
g=\phi^\ast \eta
\ee
which by construction is invariant under reparametrisations (diffeomorphisms) 
of $Z$. It is thus an example of a generally covariant field theory
and thus its canonical formulation in terms of Hamiltonian $C$ and 
spatial diffeomorphism constraints $D$ must yield a representation
of the abstract hypersurface deformation algebra of the one parameter 
family of 
hypersurfaces $t\mapsto \Sigma_t=\varphi(t,[0,1))$ dicovered in \cite{13}.
Using standard methods one finds 
\be \label{2.3}
H=P\;X'+Y\;T'+\frac{1}{2}[\Pi^2+(\Phi')^2],\;
D=P\;T'+Y\;X'+\Pi\;\Phi'
\ee
where $\dot{(.)}=\partial_t(.),\;(.)'=\partial_x(.)$ and $(P,Y,\Pi)$ are
the momenta conjugate to $(T,X,\Phi)$ respectively, i.e. the non-trivial
equal $t$ Poisson brackets are 
\be \label{2.4}
\{P(u),T(v)\}= 
\{Y(u),X(v)\}= 
\{\Pi(u),\Phi(v)\}=\delta(u,v)
\ee
with the $\delta$ distribution on $S^1$
\be \label{2.5}
\delta(u,v)=\sum_{n\in \mathbb{Z}}\; e^{i\; 2\pi\;(u-v)\;n}
\ee
One quickly verifies the hypersurface deformation algebra $\mathfrak{h}$
relations 
\be \label{2.6}
\{D(f),D(g)\}=D([f,g]),\; 
\{D(f),H(g)\}=H([f,g]),\; 
\{H(f),H(g)\}=D([f,g]);\;[f,g]:=f'\;g-f\;g'
\ee
where $f,g$ are periodic, real valued smearing functions on $S^1$ and e.g. 
$D(f)=\int_{S^1}\; dx\; f\; D$. Geometrically, $C,D$ are scalar densities 
of weight two, $f,g$ are scalar densities of weight minus one which is why
$[f,g]$ is independent of the spatial metric $q=g_{xx}$, an effect that can
happen only in one spatial dimension.

We note that the constraints depend only on the derivatives of $X,T,\Phi$
and thus do not contain information about their respective zero modes. 
We 
denote them by $\Phi_0,X_0,T_0$. Also, since $X$ is not periodic
in contrast to $Y,P,\Pi,T,\Phi$, $X'$ has a phase space independent 
zero mode given by $2\pi R$. We thus write 
\be \label{2.6a}
X(x)=2\pi\;R\;x+\tilde{X}(x)
\ee
where $\tilde{X}$ has the same zero mode as $X$ and is still conjugate to 
$P$. We can thus write the constraints 
as 
\be \label{2.6b}
D=2\pi R\; Y+\tilde{D},\;
H=2\pi R\; P+\tilde{H},\;
\ee
where $\tilde{D},\tilde{H}$ differ from $D,H$ upon replacing $X$ by 
$\tilde{X}$.
The zero modes of $Y,P,\Pi$ can be extracted 
as 
\be \label{2.6c}
Y_0=Q_\perp\cdot Y:=\int_0^1\;dx \; Y(x),\;Q:=1_L-Q_\perp
\ee
and similar for $P_0,\Pi_0$. Note $Q$ is an orthogonal projection on 
$L:=L_2([0,1),dx)$ extracting the non-zero modes of a function.  
 
It is convenient to introduce the field combinations
\be \label{2.7}
X_\pm:=\tilde{X}\pm T,\; P_{\pm}:=\frac{1}{2}(Y\pm P),\; 
A_\pm:=P_+ \pm X_+',\;
B_\pm:=P_- \pm X_-',\;
C_\pm:=\Pi\pm \Phi'
\ee
in terms of which we can write the constraints as 
\be \label{2.8}
D_\pm:=
\frac{1}{2}(\tilde{D}\pm \tilde{H}),\;
\tilde{D}_+=\frac{1}{4}[(A_+)^2-(A_-)^2+(C_+)^2],\; 
\tilde{D}_-=\frac{1}{4}[(B_+)^2-(B_-)^2-(C_-)^2]
\ee
One checks
\be \label{2.9}
\{A_\pm(u),A_\pm(v)\}=\pm\; 2\;\partial_v \delta(u,v)
\{A_\pm(u),A_\mp(v)\}=0
\ee
and similar for $B,C$, all other brackets vanishing, so that
\be \label{2.10}
\{D_\pm(f),D_\pm(g)\}=D_\pm([f,g]),\;\;  
\{D_\pm(f),D_\mp(g)\}=0
\ee
The original variables can be recovered from $A_\pm,\; B_\pm, C_\pm$ except 
for the zero modes of the configuration variables
\ba \label{2.11}
&& \Pi=\frac{1}{2}[C_+ + C_-],\;\;  
Y=P_+ + P_-=\frac{1}{2}[A_+ + A_- + B_+ + B_-],\;\;
P=P_+ - P_-=\frac{1}{2}[A_+ + A_- - B_+ - B_-]
\nonumber\\
&& \Phi'=\frac{1}{2} [C_+ - C_-],\;\;  
\tilde{X}'=X_+' + X_-'=\frac{1}{2}[A_+ - A_- + B_+ - B_-],\;\;
T'=X_+' - X_-'=\frac{1}{2}[A_+ - A_- - B_+ + B_-]
\nonumber\\
&&
\ea
so that the zero modes of $Y,P,\Pi$ but not those of $\tilde{X},T,\Phi$ are 
available from $A_\pm,B_\pm,C_\pm$. For the original constraints we find 
\be \label{2.12}
\tilde{D}_\pm=\frac{1}{2}[D\pm H]=D_\pm+2\pi\; R\; P_\pm
\ee
with 
$P_+=\frac{1}{2}[A_+ + A_-],\;P_-=\frac{1}{2}[B_+ + B_-]$. Therefore 
also
\be \label{2.13}
\{\tilde{D}_\pm(f),\tilde{D}_\pm(g)\}=\tilde{D}_\pm([f,g]),\;\;  
\{\tilde{D}_\pm(f),\tilde{D}_\mp(g)\}=0
\ee
In what follows we will only consider the algebra of the $D_\pm(f)$. The 
algebra of the $\tilde{D}_\pm$ can be treated by identical methods. 

\subsection{Quantum Theory}
\label{s2.2}  
   
The classical system consists of three independent scalar fields $X,T,\Phi$
which are coupled via the constraints which are only quadratic in the fields 
and their momenta. We thus use a Fock representation. In most approaches 
to PFT and also the closed bosonic string \cite{22} one constructs a Fock
space using the mode functions $e_n(x):=\exp(i\;2\pi\; n\; x)$ which form an 
orthonormal basis of the ``one particle Hilbert space'' $L=L_2([0,1),dx)$ and 
defines $A_\pm(n):=A_\pm(e_n)=\int\; dx\; e_n(x)^\ast A_\pm(x)$ etc. 
from which one finds $A_\pm(n)^\ast=A_\pm(-n)$
\be \label{2.14}
\{A_\pm(n_1),A_\pm(n_2)\}=\pm\;2\;(i n_2)\;\delta_{n_1+n_2,0}
\ee
or in terms of commutators
\be \label{2.15}
[A_\pm(n_1),A_\pm(n_2)]=\pm\;2\;n_1\;\delta_{n_1+n_2,0}
\ee
This allows to interpret $A_+(n)$ as an annihilation operator and 
$A_+(n)^\ast$ as a creation operator for $n>0$, 
$A_-(n)$ as an annihilation operator and 
$A_-(n)^\ast$ as a creation operator for $n<0$, while $A_+(0)=A_-(0)=(P_+)_0$
(zero mode).
Similar rermarks hold for $B_\pm,\;C_\pm$ where 
$B_+(0)=B_-(0)=(P_-)_0$ and 
$C_+(0)=C_-(0)=(\Pi)_0$. This split with respect to the sign of $n$ makes 
the discussion somewhat cumbersome as it requires to introduce six different 
Fock spaces and a separate discussion of the zero mode sector.

Let us therefore introduce the quantities
\be \label{2.16}
A:=\frac{1}{\sqrt{2}}[\omega^{1/2}\;Q\;X_+ -i \omega^{-1/2}\; Q\; P_+],\;\; 
B:=\frac{1}{\sqrt{2}}[\omega^{1/2}\;Q\;X_- -i \omega^{-1/2}\; Q\; P_-],\;\; 
C:=\frac{1}{\sqrt{2}}[\omega^{1/2}\;Q\;\Phi -i \omega^{-1/2}\; Q\; \Pi]
\ee
where 
\be \label{2.17}
\omega^2(.)=-(.)^{\prime\prime}=: -\Delta
\ee
is minus the Laplacian on $S^1$. The quantities (\ref{2.16}) are the standard 
annihilation operators of three massless Klein-Gordon fields where we 
have been careful to remove the zero mode on which the Laplacian is not 
invertible (if there would be a mass term, we would have $\omega^2=m^2-\Delta$
and in this case a separate discussion of the zero mode is not necessary). 

For the zero modes we set     
\ba \label{2.18}
&& A_0:=\frac{1}{\sqrt{2}}[\omega_0^{1/2}\;Q_\perp\;X_+ -i \omega_0^{-1/2}\; 
Q_\perp\; P_+],\;\; 
B_0:=\frac{1}{\sqrt{2}}[\omega_0^{1/2}\;Q_\perp\;X_- -i \omega_0^{-1/2}\; 
Q_\perp\; P_-],\;\; 
\nonumber\\
C_0 &:=&\frac{1}{\sqrt{2}}[\omega_0^{1/2}\;Q_\perp\;\Phi -i \omega_0^{-1/2}\; 
Q_\perp\; \Pi],\;\; 
\ea
where $\omega_0>0$ is an arbitrary parameter of dimension of inverse length.
It is therefore natural to set it equal to $1/R$ but we will keep it unfixed
for the moment.

For any operator valued distribution 
$O$ and and any 
smearing function $f$ we set 
\be \label{2.18a}
<f,O>:=\int\;dx\;f^\ast(x)\;O(x)=:O(f^\ast)
\ee
Then, by promoting the Poisson brackets to commutators  
\be \label{2.18b}
[<f,A_0>,<g,A_0>^\ast]=<f,Q_\perp g>,\;\;
[<f,A>,<g,A>^\ast]=<f,Q g>,\;\;
\ee
and simililar for the $B,C$ sectors, all other commutators vanishing. Here
$\ast$ is the respective complex conjugate of (\ref{2.16}), (\ref{2.17}) 
extended to an involution on linear combinations of products.

The relation among these annihilators is as follows
\ba \label{2.19}
A_\pm &=& P_+ \pm X_+'=Q_\perp P_+ + Q(P_+ \pm X_+')
\nonumber\\
&=& i\sqrt{\frac{\omega_0}{2}}[A_0-A_0^\ast]
+i\sqrt{\frac{\omega}{2}}[A-A^\ast]
\pm\sqrt{\frac{1}{2\omega}}[A+A^\ast]'
\nonumber\\
&=&
i\sqrt{\frac{\omega_0}{2}}[A_0-A_0^\ast]
+i\sqrt{2\omega}\;\{[Q_\pm A]-[Q_\pm A]^\ast\}
\ea
where 
\be \label{2.20}
Q_\pm=\frac{1}{2}\;[1_L\mp i \frac{\partial}{\omega}]\;Q
\ee
projects onto the positive/negative Fourier modes: $Q_\pm e_n=e_n$ if $n>/<0$ 
and zero otherwise. Note that $Q_\pm$ is an orthogonal (i.e. self-adjoint)
projection 
on the 1-particle Hilbert space $L$ which commutes with $Q,\partial,\omega$
which can be seen by using the common eigenbasis $e_n$.
As $[Q_\pm A]^\dagger=Q_\mp A^\dagger$ it follows
\be \label{2.21}
i\sqrt{2\omega} A=Q_+ A_+ + Q_- A_-
\ee
which demonstrates that the Fock space defined by declaring $A$ as 
annihilation operators is the same as the tensor product of Fock spaces 
defined by declaring $Q_+ A_+,Q_- A_-$ as annihilators which is exactly 
relation (\ref{2.15}). Similar statements hold for the $B,C$ sectors.
It is thus equivalent but more economic to work with $A$ rather than $A_\pm$
and we consider the Fock space $\cal H$ with Fock vacuum $\Omega$ annihilated
by $A_0,A$. 

We compute the commutators corresponding to (\ref{2.10}). We intrdocuce the 
building blocks
\be \label{2.22}  
E_0:=\sqrt{\omega_0}[A_0-A_0^\ast],\;\;
E_\pm:=\sqrt{\omega}\;Q_\pm\; A
\ee
so that
\be \label{2.23}
A_\pm=i(\frac{1}{\sqrt{2}}\;E_0+\sqrt{2}[E_\pm-E_\pm^\ast])
\ee
Since we need $A_\pm^2$, there is an ordering ambiguity w.r.t. the term
$(E_\pm-E_\pm^\dagger)^2$. We pick normal ordering wrt the annihilators 
$A$ and leave a possible normal ordering constant proportional to the 
algebraic 
unit $1$ open for the moment, that is we set 
\be \label{2.24}
A_\pm^2(f)=-[\frac{1}{2}\;E_0^2(Q_\perp f)+2\; E_0(1)\;
(E_\pm(f)-E_\pm(f)^\ast)(f)+2\;:(E_\pm-E_\pm^\ast)^2:(f)]=:T^0_\pm(f)+
T^1_\pm(f)+T^2_\pm(f)
\ee
where $:(.):$ denotes normal ordering. We have used 
in (\ref{2.24}) that $f$ is real valued. As $[E_0,E_\pm]=0$ we find with 
$s,s'=\pm$
\be \label{2.25}
[A_s^2(f),\;A_{s'}^2(g)]
=[T^1_s(f),T^1_{s'}(g)]
+[T^1_s(f),T^2_{s'}(g)]-[T^1_{s'}(g),T^2_s(f)]
+[T^2_s(f),T^2_{s'}(g)]
\ee
We have with 
\be \label{2.26}
E_s(f)=<\omega^{1/2}\;Q_s\;f,A>,\;
\ee
that
\ba \label{2.26}
&&
[T^1_s(f),T^1_{s'}(g)]=
4\; E_0(1)^2
\;[E_s(f)-E_s(f)^\ast,E_{s'}(g)-E_{s'}(g)^\ast]
\nonumber\\
&=&
-4\; E_0(1)^2\;\{
[E_s(f),E_{s'}(g)^\ast]
-[E_{s'}(g),E_s(f)^\ast]\}
\nonumber\\
&=&
-4\; E_0(1)^2\;\{
<\omega^{1/2}\;Q_s\;f,\omega^{1/2}\;Q_{s'}\;g>
-
<\omega^{1/2}\;Q_{s'}\;g,\omega^{1/2}\;Q_s\;f>
\}
\nonumber\\
&=&
-4\;\delta_{ss'}\; E_0(1)^2\;\{
<f,\omega;Q_s\;g>
-<g,\omega\;Q_s\;g>
\}
\nonumber\\
&=&
-2\;\delta_{ss'}\; E_0(1)^2\;(-i s)\{<f,g'>-<g,f'>\}
\nonumber\\
&=& 4\;i\;s\;\delta_{ss'}\; T^0_s([f,g])
\ea
where we used that $2\;\omega\; Q_s=1-i s\partial$. Next
\ba \label{2.27}
&& [T^1_s(f),T^2_{s'}(g)]=
4\;E_0(1)\;[E_s(f)-E_s(f)^\ast,[E_{s'}]^2(g)+[E_{s'}^\ast]^2(g)
-2[E_{s'}^\ast \; E_{s'}](g)]
\nonumber\\
&=&
4\;E_0(1)\;\{
[E_s(f),[E_{s'}^\ast]^2(g)-2[E_{s'}^\ast \; E_{s'}](g)]
-
[E_s(f)^\ast,[E_{s'}]^2(g)-2[E_{s'}^\ast \; E_{s'}](g)]
\}
\nonumber\\
&=&
8\;E_0(1)\;
\int\; dx\; dy\;f(x)\;g(y)\;
[K_{ss'}(x,y)\;\{E_{s'}^\ast(y)-E_{s'}(y)\}
-K_{s's}(y,x)\;\{E_{s'}^\ast(y)-E_{s'}(y)\}
]
\ea
with the kernel
\be \label{2.28}
K_{ss'}(x,y)=[E_s(x),E_{s'}^\ast(y)]
\;\;\Rightarrow\;\;
K_{ss'}(x,y)^\ast \; 1_{{\cal H}}
=[K_{ss'}(x,y) \; 1_{{\cal H}}]^\ast=K_{s's}(y,x)
\ee
Explicitly 
\be \label{2.28a}
<f,K_{ss'}\cdot g>=[<f,E_s>,<g,E_{s'}>^\ast]=
\delta{ss'}\;<f,\omega Q_s\;g>=:\delta_{ss'}\;<f,K_s\cdot g>
\ee
Abbreviating $E^g_{s'}(y)=g(y)\; E_{s'}(y)$ we obtain 
\ba \label{2.29}
&& [T^1_s(f),T^2_{s'}(g)]=
-8\;E_0(1)\;
\{
<f,K_{ss'}\cdot \{E^g_{s'}-[E^g_{s'}]^\ast\}>
-<f,K_{ss'}^\ast\cdot \{E^g_{s'}-[E^g_{s'}]^\ast\}>
\}
\nonumber\\
&=&
-8\;E_0(1)\;
<[K_{ss'}-K_{ss'}^\ast]\cdot f, \{E^g_{s'}-[E^g_{s'}]^\ast\}>
\}
\nonumber\\
&=& -8\;E_0(1)\;\delta_{ss'}
\{
<f,\omega\;(Q_s-Q_{-s})\{E^g_s-[E^g_s]^\ast\}>
\}
\nonumber\\
&=& 8\;E_0(1)\;\delta_{ss'} (i\;s)\;
\{
<f,\{E^g_s-[E^g_s]^\ast\}'>
\}
\ea
whence
\ba \label{2.30}
&& [T^1_s(f),T^2_{s'}(g)] - [T^1_{s'}(g),T^2_s(f)]=
= 8\;E_0(1)\;\delta_{ss'}\; (i s)
\{
<f,\{E^g_s-[E^g_s]^\ast\}'>
-<g,\{E^f_s-[E^f_{s'}]^\ast\}'>\}
\nonumber\\
&=& -8\;E_0(1)\;\delta_{ss'}\; (- i s)
\;[E_s-(E_s)^\ast](f\; g'-f'\;g)
\nonumber\\
&=& 4\; i\; s\delta_{ss'}\;T^1_s([f,g])
\ea
Finally
\ba \label{2.31}
&& [T^2_s(f),T^2_{s'}(g)]
=4\;
\int\; dx\; \int\; dy\; f(x)\; g(y)\;\times
\nonumber\\
&& 
\{
[E_s(x)^2, [E_{s'}(y)^\dagger]^2-2\; E_{s'}^\ast(y)\; E_{s'}(y)]
+
[E_s(x)^\ast]^2, [E_{s'}(y)]^2-2\; E_{s'}^\ast(y)\; E_{s'}(y)]
\nonumber\\
&& -2\; 
[E_s(x)^\ast\; E_s(x), 
[E_{s'}(y)^\ast]^2+
[E_{s'}(y)]^2-2\; E_{s'}^\ast(y)\; E_{s'}(y)]
\}
\nonumber\\
&=&
4\;\int\; dx\; \int\; dy\; f(x)\; g(y)\;\times
\nonumber\\
&& 
\{
K_{ss'}(x,y)\;[
2\;E_s(x)\;E_{s'}(y)^\ast
+2\;E_{s'}(y)^\ast\; E_s(x)
-4\; E_s(x)\;E_{s'}(y)
]
\nonumber\\
&& 
-K_{s's}(y,x)\;[
2\;E_s(x)^\ast\;E_{s'}(y)
2\;E_{s'}(y)\; E_s(x)^\ast
-4\; E_s(x)^\dagger\;E_{s'}(y)^\ast
]
\nonumber\\
&& -2(
K_{ss'}(x,y)\;E_s(x)^\dagger[2\; E_{s'}(y)^\ast-2 E_{s'}(y)]
-K_{s's}(x,y)\;[2\; E_{s'}(y)-2\;E_{s'}(y)^\ast]\; E_s(x)
)
\}
\nonumber\\
&=&
4\;\int\; dx\; \int\; dy\; f(x)\; g(y)\;\times
\nonumber\\
&& 
\{
K_{ss'}(x,y)\;
[
2\;K_{ss'}(x,y)
+4\;[E_{s'}(y)^\ast  - E_{s'}(y)]\; E_s(x)
]
-
K_{s's}(y,x)\;
[
2\;K_{s's}(y,x)
-4\;E_s(x)^\ast\;[E_{s'}(y)^\ast-E_{s'}(y)]
]
\nonumber\\
&& -4(
K_{ss'}(x,y)\;
E_s(x)^\ast\;[E_{s'}(y)^\ast-E_{s'}(y)]
-
K_{s's}(y,x)\;
[E_{s'}(y)-E_{s'}(y)^\ast]\; E_s(x)
)
\}
\ea
Since $K_{ss'}=\delta_{ss'} K_s$ we can simplify (\ref{5.31})
using $F_s(x):=E_s(x)-E_s(x)^\dagger$ and $f_x=f(x),\;g_y=g(y)$
\ba \label{2.33}
&& [T^2_s(f),T^2_{s'}(g)]
=8\;\delta_{ss'}\;\int\; dx\; \int\; dy\;f(x)\; g(y)\;\times\;
\{
K_s(x,y)\;
[
K_s(x,y)
-2\;F_s(y)]\; E_s(x)
]
\nonumber\\
&& -
K_s(y,x)\;
[
K_s(y,x)
+2\;E_s(x)^\ast\;F_s(y)
]
\nonumber\\
&& +2\; K_s(x,y)\; E_s(x)^\ast\; F_s(y)
+2\; K_s(y,x)\; F_s(y)\;E_s(x)
\}
\nonumber\\
&=& 8\;\delta_{ss'}\;\int\; dx\; \int\; dy\;K_s(x,y)\;\times\;
\{
f_x\; g_y\;
[
K_s(x,y)
-2\;:\;E_s(x)\;\;F_s(y)\;:]
\nonumber\\
&&
+2\;:\;E_s(x)^\ast\; F_s(y)\;:
]
-f_y\; g_x\;
[
K_s(x,y)
+2\;:\; E_s(y)^\ast\; F_s(x)\;:
-2\;:\; E_s(y)\; F_s(x)\;:
]
\}
\nonumber\\
&=& 8\;\delta_{ss'}\;\int\; dx\; \int\; dy\;K_s(x,y)\;\times
\nonumber\\
&& \{
f_x\; g_y\;
[
K_s(x,y)
-2\;:\;F_s(x)\;\;F_s(y)]\;:
]
-f_y\; g_x\;
[
K_s(x,y)
-2\;:\; F_s(y)\; F_s(x)\;:
]
\nonumber\\
&=& 8\;\delta_{ss'}\;\int\; dx\; \int\; dy\;K_s(x,y)\;\times
\nonumber\\
&& \{f_x\; g_y-f_y\; g_x\}\;
\{
K_s(x,y)-2\;:\;F_s(x)\;\;F_s(y)]\;:
\}
\ea
where we used that within the normal ordering symbol operator valued 
distributions commute. Using $F^f_s(x)=f(x) \; F_s(x)$ the second term
in (\ref{2.33} can be written
\ba \label{2.35}
&&
-16\;\delta_{ss'}
\;:\;
[F^f_s(\omega\; Q_s\cdot F^g_s)
-F^g_s(\omega\; Q_s\cdot F^f_s)
]
\;:
\nonumber\\
&=&
-8\;\delta_{ss'}\; (-i\; s)
\;:\;
[F^f_s((F^g_s)')
-F^g_s((F^f_s)')
]
\;:
\nonumber\\
&=&
-8\;\delta_{ss'}\; (-i\; s)
\;:\;[F_s^2]\;:(f\; g'-f'\;g)
\nonumber\\
&=& 4\; i\;s\;\delta_{ss'}\; T^2_s([f,g])
\ea
where we used that the operator $\omega$ is symmetric.

The first term in (\ref{2.33}) can be evaluated as follows:
Let $\mathbb{Z}_s=\{n\in \mathbb{Z};\;sn\ge 0\}$ then
\be \label{2.36}
(K_s \cdot f)(x)=2\pi\;\sum_{n\in \mathbb{Z}_s}\; |n|\; e_n(x)\;<e_n,f>\;\;
\Rightarrow\;\; K_s(x,y)=2\pi\;\sum_{n\in \mathbb{Z}_s}\; |n|\; 
e_n(x)\;e_n(y)^\ast
\ee
Thus 
\ba \label{2.36}
&& \int\; dx\; \int\; dy\;
f_x\; g_y\;K_s(x,y)^2
=(2\pi)^2\;\sum_{n_1,n_2\in \mathbb{Z}_s}\;|n_1\; n_2|
<f,e_{n_1+n_2}>\;<e_{n_1+n_2},g>
\nonumber\\
&=&(2\pi)^2\;
\sum_{n_1,n_2\in \mathbb{Z}_+}\;n_1\; n_2\;
<f,e_{s(n_1+n_2)}>\;<e_{s(n_1+n_2)},g>
\nonumber\\
&=&(2\pi)^2\;
\sum_{n\in \mathbb{Z}_+}\; <f,e_{s\;n}>\;<e_{s\;},g>\;
[\sum_{n_1=0}^n\;n_1(n-n_2)]
\nonumber\\
&=&(2\pi)^2\;
\sum_{n\in \mathbb{Z}_+}\; <f,e_{s\;n}>\;<e_{s\;n},g>\;
[n\;\frac{1}{2}\;n\;(n+1)-\frac{1}{6}\;n\;(n+1)\;(2n+1)]
\nonumber\\
&=&(2\pi)^2\;\frac{1}{6}\;
\sum_{n\in \mathbb{Z}_+}\; <f,e_{s\;n}>\;<e_{s\;n},g>\;
[n^3-n]
\nonumber\\
&=&(2\pi)^2\;\frac{s}{6}\;
\sum_{n\in \mathbb{Z}_s}\; <f,e_{s\;n}>\;<e_{s\;n},g>\;
[(s\;n)^3-(s\;n)]
\nonumber\\
&=&(2\pi)^2\;\frac{s}{6}\;
\sum_{n\in \mathbb{Z}}\; <Q_s\; f,e_n>\;<e_n,g>\;[n^3-n]
\nonumber\\
&=&(2\pi)^2\;\frac{s}{6}\;
\sum_{n\in \mathbb{Z}}\; <Q_s\; f,e_n>\;<
[(\frac{-i\partial}{2\pi})^3
-\frac{-i\partial}{2\pi}]\;e_n,g>
\nonumber\\
&=&(2\pi)^2\;\frac{s}{6}\;
\sum_{n\in \mathbb{Z}}\; <Q_s\; f,e_n>\;<e_n,\;
[(\frac{-i\partial}{2\pi})^3
-\frac{-i\partial}{2\pi}]\;g>
\nonumber\\
&=&(2\pi)^2\;\frac{s}{6}\;
<Q_s\; f,\;
[(\frac{-i\partial}{2\pi})^3
-\frac{-i\partial}{2\pi}]\;g>
\nonumber\\
&=&(2\pi)^2\;\frac{i\;s}{6}\;
<Q_s\; f,\;
[\frac{1}{(2\pi)^3}\;g^{\prime\prime\prime}+
\frac{1}{2\pi}\;g']>
\ea
Thus the first term in (\ref{2.33}) can be written
\ba \label{2.37}
&& 8\;i\;s\delta_{ss'}\; \frac{1}{6}\;(2\pi)^2\;
\{
<Q_s f,[\frac{1}{(2\pi)^3}\;g^{\prime\prime\prime}-
\frac{1}{2\pi}\;g']>
-<Q_s g,[\frac{1}{(2\pi)^3}\;g^{\prime\prime\prime}-
\frac{1}{2\pi}\;g']>
\}
\nonumber\\
&=& 4\;i\;s\delta_{ss'}\; \frac{1}{6}\;[(2\pi)^2\;
\{
<f,[\frac{1}{(2\pi)^3}\;g^{\prime\prime\prime}+
\frac{1}{2\pi}\;g']>
-<g,[\frac{1}{(2\pi)^3}\;f^{\prime\prime\prime}+
\frac{1}{2\pi}\;f']>
\}
]
\nonumber\\
&=:& 4\;i\;s\;\delta_{ss'}\; c\;S(f,g)\;
\}
\ea
where the term proportional $i\;s\;\partial/\omega$ in $Q_s$ has dropped 
out as 
$\partial^2/\omega=-\omega,\;\partial^4/\omega=\omega^3$ are symmetric 
operators on $L$. The term (\ref{2.37}) displays the anomaly of the classical 
hypersurface deformation algebra or equivalently its central extension 
with central charge $c=\frac{1}{6}$ which is called the Virasoro 
algebra with that central charge. 

Altogether
\be \label{2.38}
[A_s^2(f),A_{s'}^2(g)]=4\;i\;s\;\delta_{ss'}\;\{A_s^2([f,g])+c\;S(f,g)\}
\ee
and similar for the $B,C$ sector so that
\be \label{2.39}
[D_s(f),D_{s'}(g)]=i\delta_{ss'}\;[D_s([f,g]])+3\;c\;S(f,g)]
\ee
The Lie algebraic 2-cycle $S(f,g)=-S(g,f)$ is a 2-cocycle 
\be \label{2.39}
S([f,g],h)+S([g,h],f)+S([h,f],g)=0
\ee
by construction
but no 2-coboundary, i.e. there is no linear functional $F$ on the space 
of test functions $f$ such that $S(f,g)=F([f,g])$. Thus, the $D_s(f)$ cannot 
be modified by adding $3\c\;F(f)\cdot 1$ to obtain a proper Lie algebra.

It should be noted that the result (\ref{2.39}) is purely algebraic, it just
follows from $\ast-$algebraic relations and the chosen (normal) ordering. 
It is not necessary to assume a 
Fock representation, we just used the $\ast-$algebra generated by $A_0,A$ and 
their algebraic adjoints. In order that our intended Fock representation 
defined by $A_0,A$ etc. (thereby replacing algebraic adjoint $\ast$ by 
Hilber space adjoint $\dagger$) we must therefore 
check whether the constraints (and thus their adjoints as they are 
manifestly symmetric) are densely defined.
Since $D_{s'}$ is a linear
combination of the $A_s^2,B_s^2,C_s^2$, it will be sufficient to show that 
$A_s^2(f)$ is densely defined. Since $A_s^2(f)$ is a linear combination 
of the $T^j_s(f),\;j=0,1,2$ (see \ref{2.24}) it will be sufficient to 
consider those. Consider first the action of $A_s^2(f)$ on the 
Fock vacuum
\ba \label{2.40}
||T^0_s(F)\Omega||^2 &=& 
||\frac{\omega_0}{2}\;(Q_\perp F)\; A_0^\dagger\Omega||^2
=[\frac{\omega_0}{2}\;(Q_\perp F)]^2
\nonumber\\
||T^1_s(F)\Omega||^2 &=& 
||2\sqrt{\omega_0}\;A_0^\dagger\;E_s^\dagger(F)\;\Omega||^2
=4\omega_0\;<F,\omega Q_s\;F>
\nonumber\\
||T^2_s(F)\Omega||^2 
&=& 4\; ||:(E_\pm-E_\pm^\dagger)^2:(f)\Omega||^2
=4\; <(E_\pm^\dagger)^2(F)\Omega||^2
\nonumber\\
&=& 
8\int\;dx\;dy\; F(x)\;F(y)\;K_s(x,y)^2
\nonumber\\
&=&
(2\pi)^2\;\frac{i\;s}{6}\;
<Q_s\; f,\;
[\frac{1}{(2\pi)^3}\;F^{\prime\prime\prime}+
\frac{1}{2\pi}\;F']> 
\nonumber\\
&=&
(2\pi)^2\;\frac{(i\;s)^2}{12}\;
<\omega^{-1}\;F',\;
[\frac{1}{(2\pi)^3}\;F^{\prime\prime\prime}+
\frac{1}{2\pi}\;f']> 
\nonumber\\
&=&
(2\pi)^2\;\frac{1}{12}\;
[\frac{1}{(2\pi)^3}\;<f,\omega^{-1} F^{\prime\prime\prime\prime}>
+\frac{1}{2\pi}\;<f,\omega^{-1} F^{\prime\prime}>
\nonumber\\
&=&
(2\pi)^2\;\frac{1}{12}\;
[\frac{1}{(2\pi)^3}\;<F,\omega^3 F>
-\frac{1}{2\pi}\;<F,\omega F>
\ea
where we used that for smooth, real valued, periodic functions $F$
\be \label{2.41}
<F,F^{\prime\prime\prime}>=-\frac{1}{2}\;<((F')^2)'>=0,\;
<F,F'>=-\frac{1}{2}\;<(F^2)'>=0
\ee
Note that $[\omega/(2\pi)]^3-[\omega/(2\pi)]$ has spectrum in 
$\mathbb{N}$. 

To show that the hypersurface deformation generators are indeed 
densely defined and symmetric in the chosen Fock representation we should 
check that they map Fock states into normalisable states. It is 
convenient not to work with Fock states directly but rather with 
the states  
\be \label{2.41a}
w[f]\Omega,\;w[f]=\exp(i<f,\Phi>)
\ee
for the $C$ sector and similar for the $A,B$ sector. By choosing 
$f=\sum_{k=n}^\infty\; s_n\; b_n$ for some real valued ONB of $L$ 
one can generate all Fock states from the corresponding Weyl element 
$w[f]$ by taking suitable derivatives of (\ref{2.41a}) at $s_n=0,\;n\in 
\mathbb{N}$. This shows that the $w[f]\Omega$ with $f$ real valued
span a dense subset.
A short standard calculation reveals
\be \label{2.41b}
E_s(x)\;w[f]\Omega=i\;g_s(x)\; w[f]\;\Omega;\;\;g_s(x):=[Q_s f](x)
\ee
We establish the finiteness of the constraint operators on the Fock states
only for the most difficult piece $T^2_s(F)$, the other pieces are left to the 
reader. We have 
\be \label{2.41c}
-T^2_s(F)\;w[f]\Omega=\int\;dx\; F(x)\;([E_s(x)+ig_s(x)^\ast]^\dagger)^2\; 
w[f]\;
\Omega
\ee
Thus using the creation/annihilation algebra as in (\ref{2.40}) 
and (\ref{2.41b}) a 
straightforward calculation reveals 
\be \label{2.41d}
||T^2_s(F)\;w[f]\omega||^2 
=
||w[f]\omega||^2\;
\int\; dx\; F(x)\int\; dy\;F(y)\;[2\; K_s(x,y)^2+4\; K_s(x,y)\;
g(x)\; g(y)+g(x)^2\;g(y)^2]
\ee
where $g=[Q_s+Q_{-s}]f=Qf$.\\ 
\\
We now discuss the finiteness of (\ref{2.41d}). To be sure, if $f$ is smooth,
then finiteness is immediate. therefore, with respect to the smooth and 
quasi-local wavelet like functions introduced in \cite{27} for the purpose 
of renormalisation, the following complications do
not arise. However, the particular set of functions that 
were used for renormalisation in \cite{2,3,4}
are only piecewise smooth (in fact constant)
and display finitely many discontinuities. We therefore consider 
these functions in what follows in order to pin point which convergence 
issues arise, why passing to smoother coarse graining functions to define 
the renormalisation flow is more convenient and how one can still work
with only piecewise smooth coarse graining functions using zeta 
function regularisation. Readers not interested in these issues can 
safely skip the rest of the following paragraph.\\    
\\
{\bf Zeta function regularisation}\\
\\
The first term in (\ref{2.41d}) is of course 
the vacuum contribution (\ref{2.40}) and thus independent of $f$. We already 
showed that it is finite in (\ref{2.40}) for smooth $F$. The third term can 
be estimated by $||F||_\infty^2\; ||f||_\infty^4$ where $||.||_\infty$ 
denotes 
the supremum norm. Thus it is finite even if $f$ is a discontinuous but 
bounded function on $[0,1)$. The second term is given by (up to the factor $4$)
\be \label{2.41e}
<F\; (Qf),[2\omega Q_s]\;[F\; (Qf)]>
=<F\; (Qf),Q[\omega-is\partial]Q\;[F\; (Qf)]>
\ee
If $f$ is at least $C^1$ then the piece $-is\partial$ vanishes by a smiliar 
calculation as in (\ref{2.41}). If $f$ has discontinuities but is  
periodic and together with $F$ is real valued as is the case here then this 
piece still vanishes if we define for a step function with 
$0\le a<b<1,\; x\in [0,1)$
\be \label{2.41f}
\chi_{[a,b)}'(x)=\delta(x,a)-\delta(x,b),\; 
\chi_{[a,b)}(x)=\left\{ \begin{array}{cc}
1 & a<x<b\\
\frac{1}{2} & x=a\; \wedge\; x=b\\   
0 & x<a\; \wedge \; b<x
\end{array}
\right.
\ee
The boundary values of the step function are uniquely selected by requiring 
\be \label{2.41g}
<\chi_{[a,b)},\;\chi_{[cd)}'>+    
<\chi_{[a,b)}',\;\chi_{[cd)}>=0
\ee
for all possible (namely thirteen) orderings of $a,b,c,d$. 
These values also ensure that 
the sum of step functions for a partition of $[0,1)$ equals unity at every 
point. Thus even in the case of 
discontinuities (\ref{2.41e}) simplifies to 
\ba \label{2.41h}    
&& <F\; (Qf),\;Q\omega Q\;[F\; (Qf)]>
=<[F\; (Qf)]',\;Q\omega^{-1}\;Q\;[F\; (Qf)]'>
\\
&=& <F'\; (Qf),\;Q\omega^{-1}Q\;[F'\; (Qf)]>
+2\;<(Qf)',\;F;Q\omega^{-1}\;Q\;[F\; (Qf)]>
+<F\; (Qf)',Q\;\omega^{-1}\;Q [F\; (Qf)']>
\nonumber
\ea
We have explicitly, using the spectral theorem 
\be \label{2.41i}
2\pi G:=2\pi\; Q\omega^{-1}Q\;[F'\; (Qf)]=\sum_{n=1}^\infty \; n^{-1}\; 
[e_{n}\; <e_n,F\;(Qf)>+e_{-n}\;<e_{-n},F\;(Qf)]
\ee
pointwise in $[0,1)$, thus the modulus squared of (\ref{2.41i}) can be 
estimated 
from above by the Cauchy-schwartz inequlity and using $|e_n|=1$ pointwise
\be 
[\sum_{n=1}^\infty\; n^{-2}]\;[\sum_{n\not=0}\; |<e_n,F(Qf)>|
\le \; c\; ||F(Qf)||_L^2   
\le \; c\; (||F||_\infty\; ||(Qf)||_\infty)^2   
\ee
where $c>0$ is a constant. Thus since for bounded $L_2([0,1),dx)$ 
functions we have 
$||.||_L\le ||.||_\infty$
\be \label{2.41j}
||Q\omega^{-1}Q\;[F'\; (Qf)]||_L\le 
||Q\omega^{-1}Q\;[F'\; (Qf)]||_\infty \; \le c \;
||F||_\infty\; ||(Qf)||_\infty   
\ee
which shows that the first term in (\ref{2.41h}) is finite due 
to $||F' f||<\infty$ and the CS inequlaity. The second term is also 
finite if $Qf$ has finitely many discontinuities because the contributions 
of these discontinuities to the integral involving $(Qf)'$ amounts to 
a finite linear combination of evaluations of $F\; G$ at those points and 
both functions have finite supremum norm. The only potentially troublesome 
term is the last one which involves products of $\delta$ distributions.         
We evaluate it explicitly for the case encountered in the next sections 
namely 
\be \label{2.41k}
Qf=\sum_{m=0}^M\; f(m)\; \chi_m(x),\; \chi_m(x)=\chi_{[x_m,x_{m+1})}(x)
\ee
with real valued $f(m)$ and
characteristic functions $\chi_m$ of an interval where $M<\infty$ and 
$1\equiv 0=x_0<x_1<..<x_{M-1}<1$ is a partition of $[0,1)$. 
We find 
\ba \label{2.41l}
&& <F\;(Qf)',\;Q\omega^{-1}\; [F (Qf)']>
=\sum_{m_1,m_2}\; f(m_1)\; f(m_2)\; 
\{(F\;(Q\omega^{-1}Q[\chi_{m_2}' F])(x_{m_1})
-(F\;(Q\omega^{-1}Q[\chi_{m_2}' F])(x_{m_2})\}
\nonumber\\
&=& -M^{-1}\sum_{m_1,m_2}\; [\partial_M f](m_1)\; f(m_2)\; 
\{F\;(Q\omega^{-1}Q[\chi_{m_2}' F]\}(x_{m_1})
\nonumber\\
&=& -M^{-1}\sum_{m_1,m_2}\; [\partial_M f](m_1)\; f(m_2)\; F(x_{m_1})\;
\sum_{n=1}^\infty\; \frac{1}{n}\; 
[e_n(x_{m_1})\; <e_n,\chi_{m_2}' F>
+e_{-n}(x_{m_1})\; <e_{-n},\chi_{m_2}' F>]
\nonumber\\
&=& -M^{-1}\sum_{m_1,m_2}\; [\partial_M f](m_1)\; f(m_2)\; F(x_{m_1})\;
\sum_{n=1}^\infty\; \frac{1}{n}\; 
\{e_n(x_{m_1})\; [(e_{-n}\;F)(x_{m_2})-(e_{-n}\;F)(x_{m_2+1})]+c.c.\}
\nonumber\\
&=& M^{-2}\sum_{m_1,m_2}\; [\partial_M f](m_1)\; [\partial_M f](m_2)\;
F(x_{m_1})\; F(x_{m_2})\;
\sum_{n=1}^\infty\; \frac{1}{n}\; 
\{e_n(x_{m_1}-x_{m_2})+e_{-n}(x_{m_1}-x_{m_2})
\ea
with $(\partial_M f)(m)=M[f(m+1)-f(m)]$. It is the sum over $n\in \mathbb{N}$
in (\ref{2.41l}) which is problematic. We isolate and manipulate it as follows
\ba \label{2.41m}
&& \sum_{n=1}^\infty\; \frac{1}{n}\; 
\{e_n(x_{m_1}-x_{m_2})+e_{-n}(x_{m_1}-x_{m_2})
=2\sum_{n=1}^\infty\; \frac{1}{n}\;\cos(k_M(m_1-m_2)n)
\nonumber\\
&=&
=2\sum_{l=1}^{M-1}\; \frac{1}{l}\;\cos(k_M(m_1-m_2)l)
+2\sum_{l=0}^{M-1}\;\cos(k_M(m_1-m_2)l)\;\sum_{n=1}^\infty\; \frac{1}{l+nM}
\ea
where we considered an equidistant partition, set $k_M=2\pi/M$ and and 
exploited periodicity modulo $M$. Then for any $0\le l\le M-1$ we consider 
\be \label{2.41n}
\sum_{n=1}^\infty\; \frac{1}{l+nM}
=\sum_{n=1}^\infty\; [\frac{1}{l+nM}-\frac{1}{nM}]
+\frac{1}{M}\;\lim_{N\to\infty}\; \sum_{n=1}^N\; \frac{1}{n}
\ee
The first infinite sum in (\ref{2.41n}) converges absolutely for each $l$.
The limit of the second sum marginally diverges to the simple pole (with 
residue unity) value of the Riemann
zeta function. Consider 
\be \label{2.41o}
\gamma(N,\epsilon,\delta):=
\frac{1}{2}\;\sum_{n=1}^N\; 
[\frac{1}{n^{1+\delta+\epsilon}}+\frac{1}{n^{1+\delta-\epsilon}}]
\ee
If we take the limits $\epsilon\to 0+,\;\delta\to 0+,\; N\to\infty$ 
in exactly this order, then 
we return to (\ref{2.41n}). As usual, regularisation of infinities consists
in interchanging limits that would be allowed if the sums involved would
converge absolutely. We take the limits in the order 
$N\to \infty, \delta\to 0, \epsilon\to 0$. After $N\to\infty$ we obtain 
for $\delta>\epsilon>0$ the finite result
\be \label{2.41o}
\gamma(\epsilon,\delta):=\lim_{N\to\infty}\;\gamma(N,\epsilon,\delta)
\frac{1}{2}\;[\zeta(1+\delta+\epsilon)+\zeta(1+\delta-\epsilon)] 
\ee
where $\zeta$ is the Riemann zeta function. It has an analytic extension 
to the whole complex plane except for its simple pole $z=1$. With this 
analytic extension being understood in (\ref{2.41o}) 
we can now take $\delta\to 0$
\be \label{2.41o}
\gamma(\epsilon):=\lim_{\delta\to 0}\;\gamma(\epsilon,\delta)
\frac{1}{2}\;[\zeta(1+\epsilon)+\zeta(1-\epsilon)] 
\ee
Finally we take $\epsilon\to$ which results in the principal value of 
the zeta function at unity
\be \label{2.41p}
\gamma:=\lim_{\epsilon\to 0}\;\gamma(\epsilon)=[{\rm pv}\; \zeta](1)
=\lim_{\epsilon\to}\; \frac{1}{2}\;[\zeta(1+\epsilon)+\zeta(1-\epsilon)] 
\ee
which turns out to be finite and equal to the {\it Euler-Mascheroni} 
constant \cite{28}
\be \label{2.41q}
\gamma=\lim_{N\to\infty}\;[-\ln(N)+\sum_{n=1}^N\;\frac{1}{n}]
\ee
which is numerically $0.58$ in the second decimal precision. 

This kind of 
regularisation is of course standard in conformal field theory \cite{29}.
It would not be necessary if the functions $f$ were smooth. In the smooth
case exactly the same infinite sum of $1/n$ 
would occur but the difference would be that it is multiplied by 
n-dependent coefficients that either have compact support in $n$ or 
lead to stronger decay rendering the sum absolutely convergent. Thus in
the smooth case the result of the calculation would be dominated by 
the respective and corresponding first term in (\ref{2.41m}), 
(\ref{2.41n}). Note also that the proposed regularisation can be considered
as the regularisation  
\be \label{2.41r}
\omega^{-1} \to \frac{1}{2}\;[
\omega^{1+\delta+\epsilon}+\omega^{1+\delta-\epsilon}]
\ee
with $\delta>\epsilon>0$ and then taking the limits in the order described.
This regularisation is the price to pay when working with bounded discontinuous
functions $f$ but it extracts exactly the dominating terms that would 
arise if $f$ was smooth. The motivation for using non smooth step functions 
is that they result in coarse graing maps for purposes of renormalisation 
with almost perfect properties as we will see in the next section.
In \cite{27} we introduce smooth coarse graining maps which come 
very close to those step functions, for which above 
regularistion is not necessary and for which the finite result obtained here 
after regularisation is exact. As 
these step functions are finite position resolution approximants of smooth
continuuum functions, our manipulation is physically justified. This can also 
be seen as follows: The absolute value of both terms in (\ref{2.41}) can be 
bounded from above
(after above regularisation) by $c/M$ where 
$c=\gamma+\sum_{N=1}^\infty N^{-2}$ (observing that $n\le M-1$).
If $f(m)=M<\chi^M_m,f>$ for smooth $f$ as we assume in the next section,
with $\chi^M_m$ the characteristic function of the interval $[m/M,(m+1)/M)$
then the first term in (\ref{2.41l}) converges to the smooth continuum value
$<F f', Q\omega^{-1} Q F f'>$ as $M\to \infty$ while the two remaing 
terms can be bounded by $c\;||F f'||^2/M$ which converges to zero. 
Accordingly our zeta function regularisation (only necessary for non smooth
finite resolution approximants) ensures that the continuum limit 
(taking the finite resolution regulator $M\to\infty$) agrees with the 
direct continuum result.\\
\\
\\
With this understanding,
the hypersurface generators are densely defined on the span of 
Fock states.\\
\\
{\bf Comments on the spacce of solutions to the constraints}\\
\\
For completeness, we close this section with a few remarks on the actual 
solution of 
the quantum constraints which are mostly standard. These will not be of 
any relevance for the rest of the paper and the reader not interested in 
these remarks can safely jump to the next section.\\
\\
Not even the Fock vacuum is in the kernel of any of them not to speak 
of the joint kernel. Indeed, there can be no joint zero eigenvector $v$
of all the constraints except the zero vector due to the anomaly
\be \label{2.41}
0=[D_s(f),D_s(g)]\;v=is\;3\c\;S(f,g)\;v
\ee
In solving the constraints, we thus look not for joint zero eigenvectors 
(zero is not in the joint point spectrum) but 
for generalised joint eigenvectors (distibutions), i.e. linear functionals $l$
on a dense and invariant (under the action of the $D_s(f)$) domain 
$\cal D$ such that 
\be \label{2.42}   
l[D_s(F)\;v]=0\;\;\forall\; f,\;s\;, v\in {\cal D}
\ee
Note that the finite linear span of Fock states is dense but not invariant.
However, this also does not work for any such choice of domain, 
because if $\cal D$ is invariant then 
any such $l$ also satifies $l[[D_s(f),D_s(g)]\;v])=is\;c\;S(f,g)\;l[v]=0$
i.e. $l$ vanishes identically on $\cal D$. We thus resort, as it is 
common practice, to solving 
the equations $D_s(f)=0$ not in the strong operator topology but in the 
weak operator topology. That is, we look for a proper subspace 
${\cal D}\subset {\cal H}$ in the domain of the $D_s(f)$ 
such that the $D_s(f)\;{\cal D}\subset \overline{{\cal D}}_\perp$, i.e. 
the image of $\cal D$ under any $D_s(f)$ lies in the orthogonal complement
of (the completion of) $\cal D$. That is, for any $v,v'\in {\cal D}$ we 
impose for all $s,f$
\be \label{2.43}
<v,[D_s(F)-a\;<f>\;1_{{\cal H}}]\;v'>=0
\ee
where a possible normal ordering constant $a$ was introduced.
In other words, w.r.t. the split ${\cal H}=\overline{{\cal D}}\oplus
\overline{{\cal D}}_\perp$ all operators $D_s(f)$ contain no diagonal 
block corresponding to $\cal D$. A well known choice of $\cal D$ consists
in the solution to the system of equations
\be \label{2.44}
[D_s(e_n)-a\; \delta_n\; 1_{{\cal H}}]\;v=0;\;\;\forall\; s,\;n\ge 0
\ee
Since $D_s(e_n)^\dagger=D_s(e_{-n})$ it follows that 
(\ref{2.44}) implies (\ref{2.43}) for all $F$. The system (\ref{2.44}) does 
not suffer from the anomaly because for $m,n\ge 0$ 
\ba \label{2.45}
&& [D_s(e_m)-a\; \delta_n\; 1_{{\cal H}},\;
D_{s'}(e_n)-a\; \delta_n\; 1_{{\cal H}}]
=i\;\delta_{ss'}[D_s([e_m,e_n])+3\;c\;S(e_m,e_n)
\\
&=& i\;\delta_{ss'}(2\pi\;i(m-n)\;D_s(e_{m+n})-3\;i\;c\;(m^3-m-n^3+n)\;
\delta_{m+n,0}]
=-2\pi\delta_{ss'}(m-n)\;D_s(m+n)
\nonumber
\ea
as the second term only contribues for $m=n=0$ if $m,n\ge 0$ but then the 
prefactor vanishes. Thus the r.h.s. of (\ref{2.45}) is non-vanishing iff
$m+n>0$ so that the system of conditions (\ref{2.44}) is consistent. 
Of course other choices of $\cal D$ are equally valid such as imposing
(\ref{2.44}) for $n\le 0$ only for both values of $s$ or using
(\ref{2.44}) 
with $n\ge 0$ for $s=+$
and with $n\le 0$ for $s=-$.

Alternatively, to actually solve (\ref{2.44}) we could
use master constraint methods \cite{23},
i.e. we set
\be \label{2.45}
M:=\sum_{s,n\ge 0}\; m_n\; D_s(e_n)^\dagger\;D_s(e_n)
\ee
where $m_n>0$ are coefficients that decay sufficiently fast in order that
$M$ be densely defined in the Fock space. Then any solution $v$ of 
(\ref{2.44}) solves $M\;v=0$ and conversely any solution of $M\; v=0$ solves 
$<v,M\;v>=0$ and therefore (\ref{2.44}). The task is now to solve for the 
ground states of the master constraint $M$. One will look for them 
in the form 
\be \label{2.46}
v=v_{AB}\otimes v_c,\;v_{AB}\in {\cal H}_A\otimes {\cal H}_B,\;
v_C\in {\cal H}_C
\ee
where $v_C$ is any Fock state and $v_{AB}$ is to be determined in dependence
on $v_C$. In this way, the physical Hilbert space is isomorphic to 
${\cal H}_C$. 

This is of course expected as the PFT should be equivalent 
to the massless KG field on the cylinder. Indeed, the natural gauge 
fixing conditions $T=t, \; X=x$ reproduce this theory which one 
immediately arrives at using the corresponding reduced phase space 
quantisation. The actual solution of PFT is beyond the scope of the 
present work in which we are just interested in studying how the 
system behaves under renormalisation.

\section{Hamiltonian Renormalisation of Hamiltonian systems}
\label{s3}

This section is to recall the essential elements from \cite{4,11} to which 
the reader is referred to for more information.\\
\\
We introduce some coordinate $x\in[0,1)$ and equidistant lattices $\Lambda_M$
on $[0,1)$ with $M$ points $x_m=\frac{m}{M},\;m\in \mathbb{Z}_M:=\{0,1,2,..,
M-1\}$. Among the numbers $M\in \mathbb{N}$ we introduce the 
relation $M<M'$ iff $\frac{M'}{M}\in \mathbb{N}$ which means that $\Lambda_M$
is a sublattice of $\Lambda_{M'}$. It is not difficult to see that this 
defines a partial order and that $\mathbb{N}$ is directed with respect to 
it. 

The space of complex valued sequences $\{f_M(m)\}_{m\in \mathbb{Z}_M}$ is 
denoted by $L_M$ and given a Hilbert space structure by 
\be \label{3.1}
<f_M,f'_M>_{L_M}:=\frac{1}{M}\; \sum_{m\in \mathbb{Z}_M}\; f_M(m)^\ast\;
f'_M(m)
\ee
Let $\chi_{[a,b)}$ be the characteristic function of the left closed, right 
open interval $[a,b)\subset [0,1)$ and for $x\in [0,1)$
\be \label{3.2}
\chi^M_m(x):=\chi_{[\frac{m}{M},\frac{m+1}{M})}(x)
\ee
Consider the embedding (recall $L=L_2([0,1),dx)$)
\be \label{3.2}
I_M:\; L_M\to L;\; (I_M\;f_M)(x):=\sum_{m\in \mathbb{Z}_M}\;f_M(m)\;
\chi^M_m(x)  
\ee
which is in fact an isometry
\be \label{3.3}
<I_M\;f_M,\;I_M\; f'_M>_L=<f_M,f'_M>_{L_M}
\ee
and thus allows the interpretation of (\ref{3.1}) as the Riemann sum 
approximation of $<f,f'>_L$ with $f_M(m):=f(m/M),\;f'_M(m)=f'(m/M)$. 

For $M<M'$ we construct the embeddings 
\be \label{3.4}
I_{MM'}:\; L_M\to L_{M'};\;I_{MM'}:=I_{M'}^\dagger\; I_M
\ee
The operator $I_M^\dagger$ can be worked out explicitly
\be \label{3.5}
[I_M^\dagger\; f](m)=M\;<\chi^M_m,f>_L 
\ee
It is also an isometry
\be \label{3.6}
<I_{MM'} f_M,\;I_{MM'} f'_M>_{L_{M'}}=
<f_M,\;f'_M>_{L_M}
\ee
and these embeddings automatically obey the consistency conditions for 
all $M<M'<M^{\prime\prime}$
\be \label{3.7}
I_{M' M^{\prime\prime}}\circ I_{M M'}=I_{M M^{\prime\prime}}
\ee
This follows from the identity 
\be \label{3.8}
I_{M'}\;I_{M'}^\dagger\; I_M=I_M
\ee
which in turn is
due to the property of the $\chi^M_m$ to define partitions of $[0,1)$ which 
are nested for $M<M'$, that is
\be \label{3.9}
\chi^M_m=\sum_{l=0}^{k-1}\; \chi^{M'}_{km+l},\;\; k=\frac{M'}{M}
\ee
We can also work out $I_{MM'}$ explicitly ($k:=M'/M$)
\ba \label{3.9a}
&& (I_{MM'} f_M)(m')
=M'\;<\chi^M_{m'},I_M f_M>_L
=M'\sum_{m\in \mathbb{Z}_M}\; f_M(m)\;<\chi^M_{m'},\chi^M_m>_L=
\nonumber\\
&=& M'\sum_{m\in \mathbb{Z}_M}\; f_M(m)\;
(\frac{1}{M'}\sum_{l=0}^{k-1}\;\delta_{m',mk+l})
=f_M([M\frac{m'}{M'}])
\ea
where $[.]$ denotes the floor function (Gauss bracket).

Consider a scalar field $\phi$ on $[0,1)$ with conjugate momentum $\pi$.
Note that geometrically $\pi$ is a scalar density of weight one on  $[0,1)$ 
as one can see from the Poisson bracket 
\be \label{3.10}
\{\pi(x),\phi(y)\}=\delta(x,y)
\ee
We consider real density one valued test functions $f$ and 
real density 
zero valued test functions $F$ on $[0,1)$. Then the real numbers
\be \label{3.11}
\phi(f):=<f,\phi>,\;
\pi(F):=<F,\pi>
\ee
are invariant under diffeomorphisms of $[0,1)$ and we have 
\be \label{3.12}
\{\pi(F),\phi(f)\}=F(f)
\ee
One can construct the abstract $^\ast-$algebra (even $C^\ast-$algebra) 
$\mathfrak{A}$ 
generated by the Weyl elements
\be \label{3.13}
w(f,F)=\exp(i[\phi(f)+\pi(F)])
\ee
and the corresponding Weyl relations that follow from the reality of 
(\ref{3.11}) and (\ref{3.12}).  

Representations of $\mathfrak{A}$ can be constructed from a state (positive,
normalised, linear functional) 
$\hat{\omega}$ on it via the GNS construction \cite{18}. This delivers 
a Hilbert space $\cal H$, a representation $\rho$ of $\mathfrak{A}$ 
by bounded operators on 
$\cal H$ and a vector $\Omega\in {\cal A}$ cyclic for $\rho(\mathfrak{A})$.
If $\cal H$ is separable, we always 
find an Abelian sub-$^\ast-$algebra $\mathfrak{B}$ of $\mathfrak{A}$
for which $\Omega$ is still cyclic. For instance, we can pick an ONB $e_I,\;
I\in \mathbb{Z}$ with $b_0:=\Omega$ of $\cal H$ and consider the Abelian 
group of unitary 
operators $U_I,\;I\in\mathbb{Z},\;U_I^\dagger=U_{-I}$ 
such that $U_I\; e_J=e_{I+J}$. Then
we find $a_I\in \mathfrak{A}$ such that $\rho(a_I)=U_I$ and $\mathfrak{B}$ 
is generated by those $a_I$. See \cite{11} for more details and more general 
cases. Of course the $a_I$ may in general be a very complicated (in general
infinite) linear combinations of the Weyl elements (\ref{3.13}). Still
it follows that $\cal H$ can be thought of as $L_2(\Delta(\mathfrak{B}),d\nu)$
where $\Delta(\mathfrak{B})$ is the Gel'fand spectrum (space of ``characters''
i.e.
homomorphisms $\chi:\; \mathfrak{B}\to\mathbb{C}$ equipped with the 
Gel'fand topology) of $\mathfrak{B}$ and 
$\nu$ a probability measure thereon. More precisely, there is a unitary map
$U:\;{\cal H}\to L_2(\Delta(\mathfrak{B}),d\nu)$ with 
$[U\rho(b)\Omega](\chi):=\hat{b}(\chi):=\chi(b)$ which is essentially the 
Gel'fand isomorphism.

We will assume that $\mathfrak{B}$ can be generated by the $w(f):=w(f,F=0)$
so that we can identify the space of characters with the space of 
fields $\phi$ and $\nu$ as a probability measure on that space. Indeed 
this is the case in Fock representations $\hat{\omega}=<.,.>_{{\cal H}}$
in which $w(f)\Omega$ is essentially
$\exp(i<[2\omega]^{-1/2}\;f,A>_L^\dagger)\Omega$ up to a phase where 
$A=(\sqrt{\omega}\phi-i\sqrt{\omega}^{-1}\pi)/\sqrt{2}$ is the annihilator.
Thus arbitrary linear combinations of Fock states 
$<f_1,A>^\dagger..<f_n,A>^\dagger\Omega$ can be obtained by taking 
derivatives at $s_1=..=s_n=0$ of 
$w(\sum_k \; s_k\; [2\omega]^{1/2} f_k)\Omega$ establishing that the 
span of the $w(f)\Omega$ is dense. Then $\nu$ is the Gaussian measure 
with covariance $1/(2\omega)$
\be  \label{3.14}
\nu(w(f)):=<\Omega,\;w(f)\Omega>_{{\cal H}}
=\exp(-\frac{1}{4}\;<f,\omega^{-1} f>_L)   
\ee
   
Given the injections $I_M;\; L_M\to L$ we may restrict $\phi$ to 
the subspace $I_M\;L_M$ i.e.
we define a scalar field $\phi_M$ on the lattice $\Lambda_M$ by 
\be \label{3.15}
\phi_M(f_M):=\phi(I_M\;f_M)\;\;\Rightarrow\;\; \phi_M=I_M^\dagger\phi
\ee
which provides a natural ``discretisation''. Here $\phi(f):=<f,\phi>$
for real valued $f$. As $(I_M^\dagger \phi)(m)=M<\chi^M_m,\phi>$ approaches 
$\phi(x)$ in the limit $M\to \infty$ for $m=xM$ we see that the density 
zero valued $\phi$ is smeared against the density one valued 
discretised $\delta$ 
distribution $M\chi^M$
which is diffeomorphism covariant. We may likewise define 
a discretised momentum $\pi_M=M^{-1}\;
I_M^\dagger \pi=<\chi^M_{.},\pi>$ which smears the density one valued $\pi$
against the density zero valued $\chi^M$ which is also covariant. 
Together this ensures that $\phi_M,\pi_M$
are conjugate on $\Lambda_M$
\be \label{3.16}
\{\pi_M(m),\phi_M(m')\}=M\; 
\{\pi(\chi^M_m),\phi(\chi^M_{m'})\}=M<\chi^M_m,\chi^M_{m'}>_L=\delta_{m,m'}
\ee
Although this is geometrically more natural, we will instead use 
\be \label{3.16a}
\pi_M(m):=[I_M^\dagger\pi](m),\;\;\;
\{\pi_M(m),\phi_M(m')\}=M\delta_{m,m'}
\ee
so that $\phi_M,\pi_M$ are conjugate not in the sense of a Kronecker 
$\delta$ but rather a discrete $\delta$ distribution.

Given a function $H[\phi,\pi]$ on the continuum phase space coordinatised
by the variables $\phi,\pi$ we may try to define a discretised function
\be \label{3.17}
H_M[\phi_M,\pi_M]:=H[I_M\;\phi_M,\;I_M\;\pi_M]
\ee
where the approximation $I_M I_M^\dagger \to 1_L$ as $M\to \infty$ was used. 
This indeed works as long as $H$ depends on $\pi,\phi$ only algebraically.
However, when derivatives are involved, the simple prescription (\ref{3.17})
may cause trouble because the functions $\chi^M$ are not differentiable.
This can be improved by passing to alternative, smoother 
coarse graining maps $I_M$ \cite{27} which
lead to coarse graining maps $I_{MM'}$ satisfying the consistency conditions 
(\ref{3.7}) which are essential for the renormalisation scheme. 
For the examples discussed in 
\cite{27} it turns out that the {\it natural discretisation} 
$\partial_M:= I_M^\dagger \partial I_M$ is a well defined and antisymmetric 
discrete derivative operator on $L_M$. 

To keep the presentation simple and to see into which problems one may run 
using step functions, we take the usual point of view that 
the prescription (\ref{3.17}) is as good as any other as long as 
$H_M[\phi_M,\pi_M]$ converges to $H[\phi,\pi]$ in the continuum 
limit $M\to\infty$. Noting that $\phi_M(m)=M\; <\chi^M_m,\phi>$ approaches
$\phi(x)$ as $m,M\to \infty$ if wee keep $x=m/M$ fixed we may therefore
discretise e.g. $\phi'(x)$ by 
\be \label{3.18}
I_M\; (\partial_M\;\phi_M)
\ee
where 
\be \label{3.19}
[\partial_M f_M](m):=\frac{M}{2}\;[f_M(m+1)-f_M(m-1)]
\ee
is the anti-symmetric, next neighbour, first order lattice derivative.
There are an infinite number of prescriptions such as (\ref{3.19}) which 
have the correct continuum limit in the sense mentioned above and 
therefore using any such prescription introduces a {\it discretisation 
ambiguity} into the functions $H_M[\phi_M,\pi_M]$. This ambiguity 
is drastically reduced if one uses the natural discretisation using 
smoother functions $\chi^M$ with all the desired properties as indicated
above.

Given a continuum measure $\nu$ we may construct a family of measures 
$\nu_M$ by
\be \label{3.20}
\nu_M(w_M[f_M]):=\nu(w[I_M f_M]),\;w_M[f_M]=\exp(i<f_M,\phi_M>_{L_M})
\ee
which are automatically {\it cylindrically consistent}, i.e. for all
$M<M'$
\be \label{3.21}
\nu_{M'}(w_{M'}[I_{MM'} f_M])=
\nu_M(w_M[f_M])
\ee
i.e. integrating the excess degrees of freedom in artificially writing 
the function $w_M$ of $\phi_M$ as the function $w_{M'}$ of $\phi_{M'}$ which
however depends on $\phi_{M'}$ only in terms of the {\it blocked} variables
$I_{M M'}^\dagger \phi_{M'}$ does not change the result. Conversely, under 
relatively mild technical assumptions \cite{2}, 
a cylindrically consistent family 
of measures $\nu_M$ on quantum configuration spaces 
$K_M$ can be extended to a measure $\nu$   
on a space $K$ called the {\it projective limit} of the $K_M$.
In that sense, a cylindrically consistent family is as good as the 
continuum definition but the practical advantage of the family is that 
the $\nu_M$ are easier to compute.

Consider the Hilbert spaces ${\cal H}_M=L_2(K_M,d\nu_M)$ and the embeddings
\be \label{3.22}
J_M:\;{\cal H}_M\to {\cal H}=L_2(\Phi,d\nu);\;\;
w_M[f_M]\;\Omega_M\mapsto w[I_M f_M] \Omega
\ee
which by construction are isometries. Here 
$\nu_M(.)=<\Omega_M,.\Omega_M>_{{\cal H}_M}$ and  
$\nu(.)=<\Omega,.\Omega>_{{\cal H}}$. It is also not difficult to see that 
the $J_M$ inherit from the $I_M$ the consistency properties 
\be \label{3.23}
J_{M' M^{\prime\prime}}\;  
J_{MM'}=J_{M M^{\prime\prime}}\;\;\forall\;\;M<M'<M^{\prime\prime}
\ee
where $J_{MM'}=J_{M'}^\dagger\;J_M,\;M<M'$. It follows that ${\cal H}$ is 
the inductive limit of the ${\cal H}_M$ \cite{18}. Given a symmetric quadratic 
form $H$ on $\cal H$ with dense domain $\cal D$
spanned by the $w[f]\Omega$ we may 
construct the symmetric quadratic forms $H_M:=J_M^\dagger\;H\;J_M$ which are 
automatically consistently defined: For any $M<M'$ we have 
\be \label{3.24}
J_{MM'}^\dagger\;H_{M'}\; J_{MM'}=H_M
\ee
Moreover, given $J_M \psi_M,\;J_{M'}\psi'_{M'}\in 
{\cal H}$ with 
$\psi_M\in {\cal D}_M,\;\psi'_{M'}\in {\cal D}_{M'}$ in the dense set 
of the span of vectors $w_M[f_M]\Omega_M$ etc. we find $M^{\prime\prime}>M,M'$
and can compute 
\be \label{3.25}
<J_M \psi_M,\;H\;J_{M'}\psi_{M'}>_{{\cal H}}
=<J_{M M^{\prime\prime}}\psi_M, \;H_{M^{\prime\prime}}\;
J_{M' M^{\prime\prime}}\psi'_{M'}>_{{\cal H}_{M^{\prime\prime}}}
\ee
i.e. for all practical purposes the family of quadratic forms $H_M$ is as good 
as $H$ but easier to compute. Note that $H$ is {\it not} the inductive limit
of the $H_M$ \cite{18} for two reasons: First, while $H_M$ are actually 
operators and not only quadratic forms (as the systems labelled by $M$ 
only depend on finitely many degrees of freedom) the object $H$ is in general
not. Second, for $H$ to be the inductive limit of the $H_M$ we require 
the much stronger {\it intertwiner property} $J_M\; H_M=H\;J_M$ which 
implies $H_M=J_M^\dagger \;H\;J_M$ but not vice versa.\\     
\\
The problem that one encounters in quantising a classical Hamiltonian system 
with canonical variables $\phi,\pi$ and Hamiltonian $H$ is this: Provide 
a representation $\rho$ of the $\ast-$algebra generated by the 
$\phi(f),\pi(F)$ (or the $C^\ast-$algebra generated by the $w(f,F)$) that 
supports ``the'' Hamiltonian $H$ as a self-adjoint operator. We have 
used inverted commas as this task is ill-defined as it stands: The 
classical function $H$ typically is ill-defined when naively substituting 
the classical $\phi,\pi$ by their corresponding operator valued distributions.
The strategy of constructive QFT is to come up with quantisations of the 
simpler, well-defined (since finite dimensional - 
if both UV regulator $M$ and IR regulator $R$ 
are present) discretised 
Hamiltonian systems defined by $\phi_M,\pi_M,H_M$ and then
restrict the discretisation ambiguities inherent in these systems by 
inverting the logic: the automatic consistency 
{\it properties} of discretisations descending from a continuum quantum theories 
are imposed as consistency {\it conditions}. 

That is, we start from a family of triples 
$({\cal H}^{(0)}_M,\Omega^{(0)}_M,H^{(0)}_M)$ obtained by some prescription 
and then define a sequence (``renormalisation flow'') of such triples
$({\cal H}^{(0)}_M,\Omega^{(0)}_M,H^{(0)}_M)$ by the following rules:\\
1.\\
The maps for $M<M'$
\be \label{3.26}
J^{(n)}_{MM'}\; w_M[f_M]\;\Omega^{(n+1)}_M:= 
w_{M'}[I_{MM'}\;f_M]\;\Omega^{(n)}_{M'}
\ee
are imposed to be isometries, that is the corresponding measures are defined 
by 
\be \label{3.27}
\nu^{(n+1)}_M(w_M[f_M])=
\nu^{(n)}_{M'}(w_{M'}[I_{MM'}\;f_M])
\ee
2.\\
Using these we set 
\be \label{3.28}
H^{(n+1)}_M:=J_{MM'}^\dagger\; H^{(n)}_{M'}\; J_{MM'}
\ee
The idea is then to look for fixed points $J_{MM'},\Omega_M,{\cal H}_M,\nu_M,
H_M$ of this flow for which then all consistency conditions are satisfied 
by construction and which therefore defines a continuum theory. The hope
is then that 
at fixed points all but finitely many (so called relevant parameters) of the 
free parameters that coordinatise the discretisation
ambiguities also assume fixed values, thus rendering the theory predictive. 

In practice one cannot use (\ref{3.26}), (\ref{3.27}), (\ref{3.28})
for all $M<M'$ since for $M<M_1,M_2',\;
M_1'\not=M_2'$ e.g.
the definitions (\ref{3.27}) and (\ref{3.28}) generically 
do not agree when using 
$M'=M_1'$ or $M'=M_2'$ respectively. Thus one usually picks a fixed 
$M'(M)$ satisfying $M'(M)>M$, a popular choice being $M'(M)=2M$. Then,
relying on the intution of {\it universality}, the fixed point is hoped 
for not to depend on the choice $M'(M)$, so that at the fixed point the 
consistency conditions indeed hold for all $M<M'$.  

An automatic feature of this renormalisation scheme is that for all $M$ the 
fixed point vacuum $\Omega_M$ is a ground state of the fixed point Hamiltonian
$H_M$ if this is true for the initial data $\Omega^{(0)}_M,\;H^{(0)}_M$:
This follows inductively from 
\be \label{3.29}
H^{(n+1)}_M\Omega^{(n+1)}_M
=[J^{(n)}_{M M'(M)}]^\dagger\; H^{(n)}_{M'(M)}\;J^{(n)}_{M M'(M)}\;
\Omega^{(n+1)}_M
=[J^{(n)}_{M M'(M)}]^\dagger\; H^{(n)}_{M'(M)}\;
\Omega^{(n)}_{M'(M)}
=0
\ee
This condition is necessary in order to make the renormalisation scheme 
compatible with Wilsonian renormalisation of the Euclidian (path integral)
formulation from which the present scheme was derived via Osterwalder-Schrader
(OS) reconstruction \cite{4,11}.

\section{Hamiltonian renormalisation of constrained systems}
\label{s4}

As mentioned, the scheme reviewed in the previous section was motivated 
using the Euclidian formulation of a QFT which needs as a minimal input 
a self-adjoint Hamiltonian $H$ on a Hilbert space $\cal H$ 
bounded from below with vacuum $\Omega$. From these one can attempt 
to construct the associated Gibbs measure $\mu$ on the space of field histories
and when this exists, it satisfies a minimal set of Euclidian axioms
(in particular reflection positivity) ensuring that $(H,{\cal H},\Omega)$ 
can be recovered from $\mu$.

When we consider constrained Hamiltonian systems, in particular when there 
is no Hamiltonian but just a set of Hamiltonian constraints, we are strictly
speaking leaving that framework. One can return to it by using the reduced
phase space formulation in which one gauge fixes the Hamiltonian constraints 
thereby ending up with a true Hamiltonian again which just acts on the 
gauge invariant (or true) degrees of freedom \cite{24} and this is the 
strategy followed so far \cite{11}. However, in this paper we want to explore
a different route:\\
The observation is that the two renormalisation steps (\ref{3.27}) and 
(\ref{3.28}) actually do not rely on $H$ being bounded from below 
or that $\Omega$ is the vacuum of $H$. Thus we propose to ``abuse''
(\ref{3.27}) and (\ref{3.28}) and use them also for constrained Hamiltonian
systems. In other words, we keep (\ref{3.27}) as it is and apply (\ref{3.28})
to each constraint operator separately. 

This proposal raises two immediate questions:
\begin{itemize}
\item[1.]
The classical continuum constraints are of the form $H(F)=\int\; dx\;
F(x)\;H(x)$ where $F$ is a smearing function and $H(x)$ is the Hamiltonian 
constraint density. Thus the essential difference between a true Hamiltonian
system and a constrained Hamiltonian system (apart from the fact that true 
Hamiltonian densities are typically bounded from below at least classically) 
is that 
for the true Hamiltonian the only allowed smearing function is $F=1$ while 
for the constrained case the space of smearing function is infinite 
dimensional. The question is now how $F$ should be treated when we discretise
$H(F)$. There are two extreme and equally natural points of view: 
\begin{itemize}
\item[i.]
The first is that
for each $F$ the function $H(F)$ is simply an independent object and should 
be treated just as a true Hamiltonian. That is, the function $F$ remains 
as it is, it is not discretised.
\item[ii.]
The second is that $F$ should be treated on equal footing with the phase 
space variables $\phi,\pi$ and thus should be discretised, perhaps by
the same map $I_M^\dagger$, perhaps by another. This of course 
introduces yet more discretisation ambiguities into the quantisation 
and also requires to invent a flow equation on the space of discretised 
smearing functions $F_M$ when stating (\ref{3.28}).   
\end{itemize}
Note that the second point of view is often taken for granted in lattice 
inspired approaches to constrained systems \cite{25}.  
One may think that the first point of view in fact provides a natural 
choice of disretisation of $F$ as follows:\\ 
Suppose that we actually have the continuum theory, i.e. the Hilbert space 
$\cal H$ and the constraints $H(F)$ at our disposal. Then the idea 
is to define a map $E_M:\;L\to L_M$ via the identity 
\be \label{4.1}
H_M(E_M\;F):=\sum_{m\in \mathbb{Z}_M}\;(E_M F)(m)\; H_M(m)
:=J_M^\dagger\; H(F)\;J_M
\ee
which assumes that the r.h.s. can actually be written in this local form. This 
is unfortunately already not the case even for the PFT considered here. The 
reason for this to happen is that $H$ when written in terms of polynomials
of annihilation 
and creation operators involves non-local integral kernels. While these 
do get discertised by means of $J_M$ this leads to an effective $E_M$ which
maps $L\to L_M^N$ where $N\ge 2$ is the polynomial degree. 
We will demonstrate this
explicitly below for PFT. 

This establishes that viewpoints i. and ii. are drastically
different, i.e. a map $E_M: L\to L_M$ generically 
cannot be induced via (\ref{4.1}).
Instead, according to viewpoint ii. we consider as an extra structure 
maps 
$\tilde{I}_M\;L_M\to L$ and $\tilde{I}_{MM'}:\; L_M\to L_{M'}$ and define 
\be \label{4.2}
H_M(F_M):=
\sum_{m\in \mathbb{Z}_M}\;F_M(m)\; H_M(m):=
J_M^\dagger\; H(\tilde{I}_M F_M)\; J_M
\ee
This is consistently defined 
\be \label{4.3}
H_M(F_M)=J_{MM'}^\dagger\; H_{M'}(\tilde{I}_{MM'} F_M)\; J_{MM'}
\ee
due to $J_{M'}\; J_{MM'}=J_M$ and provided that 
$\tilde{I}_{M'}\;I_{MM'}=I_M$. We may reduce the ambiguity and 
actually consider $\tilde{I}_M=I_M,\;\tilde{I}_{MM'}=I_{MM'}$, however,
this choice is inconvenient for the following reason: While we can certainly 
compute the commutator $[H_M(F_M),H_M(G_M)]$ directly which is well defined, 
one would like to
see the deviation from the continuum computation by using the identity
\ba \label{4.3}
&&[H_M(F_M),H_M(G_M)]=J_M^\dagger\{[H(I_M F_M),H(I_M G_M)]
-H(I_M F_M)\;(1_{{\cal H}}-P_M)\;H(I_M G_M)
\nonumber\\
&& +H(I_M G_M)\;(1_{{\cal H}}-P_M)\;H(I_M F_M)\} J_M^\dagger\
\ea
where we defined $P_M=J_M J_M^\dagger$ which is a projection in $\cal H$
due to the isometry of $J_M$. The first term gives the cylindrical 
projection of the continuum algebra which in our case is the Virasoro algebra.
The second and third term should vanish as $M\to\infty$ because $J_M$ becomes 
the 
identity in $\cal H$. Therefore (\ref{4.3}) appears to be an appropriate 
way to monitor how the cylindrically projected theories approach the correct 
continuum. The catch is that we know that in PFT the commutator 
$[H(I_M F_M),H(I_M G_)]$ depends on first and third derivatives of the 
$I_M F_M, I_M G_M$ which are, however, not even continuous. Accordingly,
if we want to use (\ref{4.3}) we should instead use 
$\tilde{I}_M,\;\tilde{I}_{MM'}$ which are at least $C^3$ and which share 
all the properties of $\tilde{I}_M,\;\tilde{I}_{MM'}$. Thus such maps 
constructed from wavelets \cite{21} suggest themselves, we will give 
more details below.

To summarise this part of the discussion, for the purpose of this paper
we take viewpoint i. and leave $F,G$ un-discretised and then 
with $H_M(F)=J_M^\dagger\; H(F)\; J_M$ the 
computation 
\be \label{4.4}
[H_M(F),H_M(G)]=J_M^\dagger\{[H(F),H(G)]
-H(F)\;(1_{{\cal H}}-P_M)\;H(G)
+H(G)\;(1_{{\cal H}}-P_M)\;H(F)\} J_M^\dagger\
\ee
is unproblematic. To avoid confusion note that (\ref{4.4}) is supposed to 
yield the Virasoro algebra, as $M\to \infty$, including the central term,
i.e. the anomaly as compared to the classical computation (Witt algebra)
should be present. We thus want to check that the Virasoro algebra is 
recovered without anomaly, not the Witt algebra.
\item[2.] As noted in the previous section, due to the central term in the 
Virasoro algebra, there cannot be a joint vacuum $\Omega$ for all the 
constraints
$H(F)$. This is even more the case for the $H_M(F)$ at finite resolution 
because they typically
do not close as it is plain to see from (\ref{4.4}), hence the states
$\Omega_M$ that arise at the fixed point cannot be joint vacua for the 
$H_M(F)$.

This is no obstacle for the renormalisation scheme when applied 
separately to the $H(F)$ because the $H_M(F)$ are 
operators (and not only quadratic forms)
of systems with finitely many degrees of freedom and thus  
one does not expect the usual problems in finding a domain that is typical
for QFT (infinitely many degrees of freedom) especially if $H(F)$, even when 
normal ordered, contains 
terms that are monomials made solely from creation operators. Thus we expect 
to find dense domains $D_M(F)$ for $H_M(F)$ 
and by construction $J_{MM'} D_M(F)\subset D_{M'}(F)$. However, a problem 
may occur when we compute commutators such as (\ref{4.4}) because the domains 
$D_M(F)$ may depend on $F$ and it may be the case that 
$H_M(F) \; D_M(F)\not\subset D_{M(F')}$ \cite{26}. At least it is true that 
at finite $M$ the domains are invariant $H_M(F)\;D_M(F)\subset D_M(F)$ 
because they are just finite linear combinations of monomomials (and not
infinite linear combinations as in case of $H(F)$) of creation and annihilation
operators. Thus a minimal requirement for (\ref{4.4}) to be meaningful is that 
the $H_M(F)$ have a dense,
invariant domain $D_M$ independent of $F$ and then by construction 
$J_{MM'} D_M\subset D_{M'}$.    

Since the span $D$
of the $J_M D_M$ is dense in the inductive limit $\cal H$ on 
which by construction is a form domain of $H(F)$, this then also makes the 
fixed point $H(F)$ densely defined as a quadratic form. However, this does 
not ensure that the commutators of the $H(F)$ are well defined
because matrix elements of the formal expression $H(F) H(F')$, which can 
be formally computed by invoking resolutions of the identity in  terms of an 
$ONB$ made from vectors in $D$, may diverge, which is a potential danger 
even if $H(F)$ can be promoted to an operator especially if $D$ is not 
invariant for $H(F)$. It is here where   
a joint cyclic vacuum would be very convenient to build a common dense 
operator domain upon. In absence of
it, the construction of such a domain may be very difficult, if it exists 
at all. In PFT we know that this problem does not occur, despite the 
non-existence of such a joint vacuum, as a common dense 
(but not invariant) operator domain is given explicitly 
by the span of the chosen Fock states. However, it may 
be in more complicated theories, especially if the domains depend on 
$F$ which in unfortunate cases can have non-dense intersections \cite{25}.  
\item[3.] Note that our renormalisation scheme constructs a single Hilbert 
space $\cal H$ (or measure $\nu$) but an infinite number of quadratic forms
$H(F)$ 
if a simultaneous 
fixed point of the respective flow equations exists at all. 
While the flow equations
for $\nu$ and $H(F)$ are tightly coupled, the flow equations for the various 
$H(F)$ are treated as independent for each choice of $F$. Now it could happen 
that these latter equations have several different fixed points for each 
choice of $F$ that are reached depending on the choice of initial 
discretisation $H^{(0)}_M(F)$. Then the corresponding fixed point family
$H_M(F)$ may depend rather 
dis-continuously on $F$ and thus would probably not coincide with the result
of 
{\it blocking from the continuum} $H_M(F):=J_M\; H(F)\;J_M$.    
\end{itemize}
In the next section we examine whether these issues arise in 
the Hamiltonian renormalisation of PFT.

\section{Hamiltonian renormalisation of PFT}
\label{s5}

Since the constraint operators are of the form 
\be \label{5.1}
D_+=[A_+^2-A_-^2]\otimes\;1_B\otimes 1_C+1_A\otimes 1_B\otimes C_+^2,\;\; 
D_+=1_A\otimes [B_+^2-B_-^2]\otimes\;1_C-1_A\otimes 1_B\otimes C_-^2
\ee
it will be sufficient to consider one of the sectors $A,B,C$ only, say $C$. 
Our first task is to pick initial discretisations of the 
$C^{(0)}_{\pm,M}$ and
corresponding Hilbert space measures $\nu^{(0)}_M$ on 
$K_M=\mathbb{R}^M$. As suggested by the considerations of 
section \ref{s2} we build $C^{(0)}_{\pm,M}$ out of $C^{(0)}_{0,M},\;
C^{(0)}_M$. We define in parallel to the continuum (see (\ref{2.19}), 
(\ref{2.20}))
\ba \label{5.2}
\Phi_M &:=& I_M^\dagger \Phi
\nonumber\\
\Pi_M &:=& I_M^\dagger \Pi
\nonumber\\
Q^\perp_M f_M &:=& <1,f_M>_{L_M}
\nonumber\\
Q^M &:=& 1_{L_M}-Q^M_\perp
\nonumber\\
C^{(0)}_{0,M} 
&:=& \frac{1}{\sqrt{2}}[\sqrt{\omega_0}\;Q^M_\perp\; \Phi_M
-i\frac{1}{\sqrt{\omega_0}}\;Q^M_\perp\;M\;\Pi_M]
\nonumber\\
C^{(0)}_{M} 
&:=& \frac{1}{\sqrt{2}}[\sqrt{\omega^{(0)}_M}\;Q^M \Phi_M
-i\frac{1}{\sqrt{\omega^{(0)}_M}}\;Q^M\;\Pi_M]
\nonumber\\
C^{(0)}_{s,M} &:=& 
i\sqrt{\omega_0/2}\;[C^{(0)}_{0,M}- (C^{(0)}_{0,M})^\dagger]
+i\sqrt{2\omega^{(0)}_M}\;[Q^{M(0)}_s C^{(0)}_M- (Q^{M(0)}_s\; 
C^{(0)}_M)^\dagger)]
\nonumber\\
(\omega^{(0)}_M)^2 &:=& -(\partial^{(0)}_M)^2
\nonumber\\
(\partial^{(0)}_M f_M)(m) &:=& (2M)^{-1}\;[f_M(m+1)-f_M(m-1)]
\nonumber\\
Q^{M(0)}_s &=& \frac{1}{2}\;[1_{L_M}-i s\;
\frac{\partial^{(0)}_M}{\omega^{(0)}_M}]\;Q^M
\nonumber\\
D^{(0)}_{s,M} &=& :\;[C^{(0)}_{s,M}]^2\;:
\ea
Here the adjoint operation and normal ordering 
is with respect to the Fock Hilbert space structure ${\cal H}^{(0)}_M$
defined by the annihilation operators $C^{(0)}_{0,M},\;C^{(0)}_M$ with Fock
vacuum $\Omega^{(0)}_M$. Note that $Q^M_\perp,\;Q^M,\; i\partial_M,\; 
\omega_M$ are self-adjoint on $L_M$ and that $Q_\perp,\;Q^M,Q^M_s$ 
are orthogonal projections in $L_M$ with $Q^M_\perp Q^M=Q^M_+ Q^M_-=0$ and
$1_{L_M}=Q^M_\perp+Q^M,\;Q^M=Q^M_+ + Q^M_-$. 

An immediate observation is that 
\be \label{5.3}
Q^M_\perp \Phi_M=<1,\Phi_M>_{L_M}
=\frac{1}{M}\sum_m\; 1(m)\; \Phi_M(m)  
=\frac{1}{M}\sum_m\; (I_M^\dagger\Phi)(m)  
=\sum_m\; <\chi^M_m,\Phi>_L=<1,\Phi>=Q_\perp \Phi
\ee
and similarly for $Q^M_\perp \Pi_M=Q_\perp \Pi$ so that in fact
\be \label{5.4}
C^{(0)}_{0,M}=C_0
\ee
is actually the same as in the continuum in the initial discretisation. 
We will see that this property is preserved by the renormalisation flow 
so that the zero modes remain un-renormalised.

We proceed to the flow equation for the Fock measure. We have
\ba \label{5.5}
&& <f_M,\Phi_M>_{L_M}=
<Q^M_\perp\; f_M,\Phi_M>_{L_M} 
+<Q^M\; f_M,\Phi_M>_{L_M} 
\nonumber\\
&=&
<[2\omega_0]^{-1/2} \; Q^M_\perp\; f_M,\;C^{(0)}_{0,M}>_{L_M}
+<[2\omega_0]^{-1/2} \; Q^M_\perp\; f_M,\;C^{(0)}_{0,M}>_{L_M}]^\dagger)
\nonumber\\
&&
+<[2\omega^{(0)}_M]^{-1/2} \; Q^M\; f_M,\;C^{(0)}_M>_{L_M}
+<[2\omega^{(0)}_M]^{-1/2} \; Q^M\; f_M,\;C^{(0)}_M>_{L_M}]^\dagger)
\ea
Thus the initial measure family has generating functional of moments 
\ba \label{5.6}
&& \nu^{(0)}_M(w_M[f_M])
=<\Omega_M^{(0)},\exp(i<f_M,\phi_M>)\;\Omega^{(0)}_M>_{{\cal H}^{(0)}}  
\nonumber\\
&=& \exp(-\frac{1}{4}[
<Q^M_\perp\; f_M,\omega_0^{-1} \; Q^M_\perp\; f_M>_{L_M}  
+Q^M\; f_M,[\omega^{(0)}_M]^{-1} \; Q^M\; f_M>_{L_M}])
\ea
It is a family of Gaussian measures with covariances (kernels on $L_M$)
\be \label{5.7}
K^{(0)}_M=\frac{1}{2}[
Q^M_\perp \;\omega_0^{-1}\; Q^M_\perp
+Q^M \;[\omega^{(0)}_M]^{-1}\; Q^M]
\ee
This is exactly as for the 1+1 Klein Gordon field treated in the first 
reference of \cite{6} except that there we assumed a non-vanishing mass
$p$ so that the projections $Q^M_\perp,Q^M$ are not not necessary and 
the initial covariance is just $[2\omega^{(0)}_M(p)]^{-1}$ with 
$[\omega^{(0)}_M(p)]^2=[\omega^{(0)}_M]^2+p^2$. 

To study the flow of (\ref{5.7}) we 
can borrow the results of \cite{6} as follows:\\
In \cite{6} we used the spectral theorem to write 
\be \label{5.8}
[2\omega_M(p)]^{-1}=\int_{\mathbb{R}}\; \frac{dk}{2\pi}\; 
[k^2+(\omega^{(0)}_M(p))^2]^{-1}
\ee
by the residue theorem where due to $p\not=0$ there is no real pole of 
the holomorphic integrand. Here, instead of integrating over the real line,   
we consider the path 
\be \label{5.9}
c_\rho:\; \mathbb{R}\to\mathbb{C};\;
c_\rho(k)=\left\{ \begin{array}{cc}
k & |k|>\rho \\
-\rho\; e^{i\frac{\pi}{2}(\frac{k}{\rho}+1)}
& |k|=\rho
\end{array}
\right.
\ee
where $\rho>0$ is arbitrarily small thus avoiding the real pole $k=0$. Then 
\be \label{5.10}
Q^M\; [2\omega^{(0)}_M]^{-1}\;Q^M = \lim_{\rho\to 0+}\; \int_{c_\rho}\; 
\frac{dk}{2\pi}\; [k^2+(\omega^{(0)}_M)^2]^{-1}
\ee
By the flow equation 
\be \label{5.11}
\nu^{(n+1)}_M(w_M(f_M)):=
\nu^{(n)}_{M'(M)}(w_{M'(M)}(I_{M M'(M)}\;f_M))
\ee
the measure family stays always inside the Gaussian class and (\ref{5.12})
translates into a flow of covariances 
\be \label{5.12}
K^{(n+1)}=I_{M M'(M)}^\dagger\; K^{(n)}_{M'(M)}\; I_{M M'(M)}
\ee
where $M'(M)>M$ is the fixed higher resolution that enters the concrete 
implementation of the blocking equations. As in \cite{6} we will choose
$M'(M)=2M$ for simplicity.
 
We note that 
\ba \label{5.13}
&& Q^{M'}_\perp I_{M M'} f_M=
<1,I_{M M'} f_M>_{L_{M'}}
=\frac{1}{M'}\; \sum_{m'\in \mathbb{Z}_{M'}}\; f_M([M\;\frac{m'}{M'}])  
\nonumber\\
&=& \frac{1}{M'}\; \sum_{m\in \mathbb{Z}_M}\; f_M(m)
[\sum_{l=0}^{M'/M-1}\; 1]
=\frac{1}{M}\; \sum_{m\in \mathbb{Z}_M}\; f_M(m)=<1,f_M>_{L_M}
\nonumber\\
&=&Q^M_\perp f_M
=I_{MM'} Q^M_\perp f_M
\ea
where in the last step we used that $I_{MM'} c=c$ if $c$ is a constant.
Thus 
\be \label{5.14}
Q^{M'}_\perp\;I_{MM'}=I_{MM'}\; Q^M_\perp
\ee
i.e. the family of projections $Q^M_\perp$ is equivariant w.r.t. the 
coarse graining maps $I_{MM'}$. Similarly
\be \label{5.15}
Q^{M'} I_{MM'}
=(1_{L_{M'}}-Q^{M'}_\perp)\;I_{MM'}  
=I_{MM'}-I_{MM'}\;Q^M_\perp  
=I_{MM'}\;(1_{L_M}-Q^M_\perp)  
=I_{MM'}\;Q^M
\ee
It follows from (\ref{5.7}) and (\ref{5.12}) that the covariance always takes 
the form 
\be \label{5.16}
K^{(n)}_M=\frac{1}{2}[
Q^M_\perp \;[\omega^{(n)}_{0,M}]^{-1}\; Q^M_\perp
+Q^M \;[\omega^{(n)}_M]^{-1}\; Q^M]
\ee
in particular the projections $Q^M,Q^M_\perp$ are not changed under the flow. 
Moreover we have separated the flow
\be \label{5.17} 
[\omega^{(n+1)}_{0,M}]^{-1}=
I_{MM'(M)}^\dagger\; [\omega^{(n)}_{0,M'(M)}]^{-1}\;I_{MM'(M)},\;\;
[\omega^{(n+1)}_{M}]^{-1}=
I_{MM'(M)}^\dagger\; [\omega^{(n)}_{M'(M)}]^{-1}\;I_{MM'(M)},\;\;
\ee

The obvious fixed point of the first equation in (\ref{5.17}) is 
\be \label{5.18}
[\omega^{(n)}_{0,M}]^{-1}=\omega_0^{-1}\; Q^M_\perp
\ee
i.e. the zero modes remain  unrenormalised as promised. As for the 
second equation, we can in view of (\ref{5.10}) immediately copy the 
results of \cite{6}: Instead of the parameter $q^2:=k^2+p^2$ used there
we just use $q^2=k^2$. All other relations remain {\it literally identical}.
As the flow equations in \cite{6} depend analytically on $q^2$ we
infer that the fixed point covariance $\omega_M$ 
is the same as in \cite{6} except 
that $p=0$ and that it appears sandwiched between $Q^M$
\be \label{5.19}
K_M=\frac{1}{2}[
Q^M_\perp \;\omega_0^{-1}\; Q^M_\perp
+Q^M \;\omega_M^{-1}\; Q^M]
\ee
and moreover $K_M$ agrees with the covariance obtained by blocking from
the continuum.\\
\\
Next we turn to the smeared constraints. 
Here we enter new territory as compared
to \cite{6}, first due to the presence of the projections $Q^{M(0)}_s$ and 
second because the constraints do not annihilate the Fock vacuum.  
We focus just on the part 
of $D_s(f)$ quadratic in the non-zero mode fields as this term by itself also 
satisfies  
the Viarasoro algebra, see section \ref{s2} where this term was denoted by 
$T^2_s(f)$, and it is also this term alone which leads
to the anomaly. The other terms denoted $T^0_s(f),T^1_s(f)$ can be treated 
by similar methods. We 
start with the continuum expression and write it in terms of integral 
kernels
\be \label{5.20} 
D_s(F)=\int\; dx\; F(x)\;\int\; dy\;\int\; dz\; 
[\kappa^1_s(x;y,z) C(y)^\dagger\; C(z)
+\kappa^2_s(x;y,z) C(y)\; C(z)
+\kappa^2_s(x;y,z)^\ast C(y)^\dagger\; C(z)^\dagger]
\ee
where $\kappa^1_s(x;y,z)^\ast=\kappa^1_s(x;z,y)$ and   
$\kappa^2_s(x;y,z)=\kappa^2_s(x;z,y)$. We block from the continuum 
and compute $[D_s(f)]_M:=J_M^\dagger\; D_s(f)\;J_M$
\be \label{5.21}
<w_M[f_M]\Omega_M,[\;D_s(F)]_M\; w_M[g_M]\Omega_M>_{{\cal H}_M}
= <w[I_M\; f_M]\Omega,D_s(F)\; w[I_M\;g_M]\Omega>_{{\cal H}}
\ee
We have for any $f,g$
\ba \label{5.22}
&& <w[f]\Omega,D_s(F)\; w[g]\Omega>_{{\cal H}}
=\int\; dx\; F(x)\;\int\; dy\;\int\; dz\; 
[\kappa^1_s(x;y,z) \;<C(y\;w[f]\Omega, C(z)\; w[g]\Omega>
\nonumber\\
&& +\kappa^2_s(x;y,z) \;<w[f]\Omega, C(y)\; C(z)\; w[g]\Omega>
+\kappa^2_s(x;y,z)^\ast \; <C(y)\; C(z)\; w[f]\Omega, \; w[g]\Omega>
\ea
and   
\ba \label{5.23}
&& C(y)\; w[f]\;\Omega=
w[f]\; w[f]^{-1}\; C(y)\; w[f]\Omega=
w[f]\; (C(x)-i[\phi(f),C(y)])\;\Omega=
i[C(x),\phi(f)]\;w[f]\;\Omega
\nonumber\\
&=& [(2\omega)^{-1/2} Q f](y)\;w[f]\;\Omega
\\
&& C(y)\;C(z)\; w[f]\;\Omega=
[(2\omega)^{-1/2} Q f](z)\;
C(y)\;w[f]\;\Omega=
[(2\omega)^{-1/2} Q f](z)\;
[(2\omega)^{-1/2} Q f](y)\;w[f]\;\Omega
\nonumber
\ea
Abbreviating $\sigma=(2\omega)^{-1/2} Q$ we thus find 
\ba \label{5.24}
&& <w[f]\Omega,D_s(F)\; w[g]\Omega>
=<w[f]\Omega,\;w[g]\Omega>\;\times\;
\nonumber\\
&& \int\; dx\; F(x)\;\int\; dy\;\int\; dz\; 
(\sigma f)(y)\;(\sigma g)(z)\;
[\kappa^1_s(x;y,z)+\kappa^2_s(x;y,z) 
+\kappa^2_s(x;y,z)^\ast] 
\ea
Applied to $f=I_M f_m,\;g=I_M\;g_M$ we obtain due to 
$J_M^\dagger J_M=1_{{\cal H}_M}$
\ba \label{5.25}
&& <w_M[f_M]\Omega_M,[\;D_s(F)]_M\; w_M[g_M]\Omega_M>
\\
&=& <w_M[f_M]\Omega_M,\;w_M[g_M]\Omega_M>\;\times\;
\nonumber\\ &&
\int\; dx\; F(x)\;\int\; dy\;\int\; dz\; 
(\sigma I_M f_M)(y)\;(\sigma I_M g_M)(z)\;
[\kappa^1_s(x;y,z)+\kappa^2_s(x;y,z) 
+\kappa^2_s(x;y,z)^\ast] 
\nonumber\\
&=& <w_M[f_M]\Omega_M,\;w_M[g_M]\Omega_M>\;\times\;
\nonumber\\ &&
\sum_{m_1,m_2\in \mathbb{Z}_M}\; f_M(m_1) \; g_M(m_2)
\int\; dx\; F(x)\;\int\; dy\;\int\; dz\; 
\sigma_M(y,m_1)\;\sigma_M(z,m_2)\;\times
\nonumber\\
&&
[\kappa^1_s(x;y,z)+\kappa^2_s(x;y,z) 
+\kappa^2_s(x;y,z)^\ast] 
\nonumber\\
&=:& <w_M[f_M]\Omega_M,\;w_M[g_M]\Omega_M>\;\times\;
\nonumber\\ &&
\sum_{m_1,m_2\in \mathbb{Z}_M}\; f_M(m_1) \; g_M(m_2)
\int\; dx\; F(x)\;[
\kappa^1_{s,M}(x;m_1,m_2)
+\kappa^2_{s,M}(x;m_1,m_2)
+\kappa^2_{s,M}(x;m_1,m_2)^\ast]
\nonumber
\ea
with $\sigma_M(x,m):=(\sigma \chi^M_m)(x)$. Now in terms of 
\be \label{5.26}
C_M=\frac{1}{\sqrt{2}}[\sqrt{\omega_M}Q^M\Phi_M-i\sqrt{\omega_M}^{-1}
Q_M \Pi_M]
\ee
where $\omega_M^{-1}$ is the fixed point covariance that we obtained from 
the flow of the measures and which annihilates $\Omega_M$. We find 
with the abbreviation $\hat{\sigma}_M=[2\omega_M]^{-1/2} Q^M$
and the Ansatz 
\ba \label{5.27}
&& [D_s(F)]_M=
\sum_{\hat{m}_1,\hat{m}_2\in \mathbb{Z}_M}\; 
\int\; dx\; F(x)\;[
\hat{\kappa}^1_{s,M}(x;\hat{m}_1,\hat{m}_2)\; C_M(\hat{m}_1)^\dagger\; 
C_M(\hat{m}_2)
\nonumber\\
&& +\hat{\kappa}^2_{s,M}(x;\hat{m}_1,\hat{m}_2)\;C_M(\hat{m}_1)\; C_M(\hat{m}_2)
+\hat{\kappa}^2_{s,M}(x;\hat{m}_1,\hat{m}_2)^\ast\;
C_M(\hat{m}_1)^\dagger\; C_M(\hat{m}_2)^\dagger]
\ea
with 
\be \label{5.28}
\hat{\kappa}^1_{s,M}(x;\hat{m}_1,\hat{m}_2)^\ast=
\hat{\kappa}^1_{s,M}(x;\hat{m}_2,\hat{m}_1),\;\;
\hat{\kappa}^2_{s,M}(x;\hat{m}_1,\hat{m}_2)=
\hat{\kappa}^2_{s,M}(x;\hat{m}_2,\hat{m}_1)
\ee
by the exacly the same calculation
\ba \label{5.29}
&& <w_M[f_M]\Omega_M,\; [D_s(F)]_M \; w_M[g_M] \Omega_M>_{{\cal H}_M}
\nonumber\\
&=&
<w_M[f_M]\Omega_M,\;w_M[g_M]\Omega_M>\;\times\;
\sum_{m_1,m_2\in \mathbb{Z}_M}\; f_M(m_1) \; g_M(m_2)
\int\; dx\; F(x)\;\sum_{\hat{m}_1,\hat{m}_2}
\nonumber\\ &&
\hat{\sigma}_M(\hat{m}_1,m_1)\;\hat{\sigma}_M(\hat{m}_2,m_2)\;
[\hat{\kappa}^1_{s,M}(x;\hat{m}_1,\hat{m}_2)+
\hat{\kappa}^2_{s,M}(x;\hat{m}_1,\hat{m}_2) +
\hat{\kappa}^2_{s,M}(x;\hat{m}_1,\hat{m}_2)^\ast] 
\nonumber\\
&=:& <w_M[f_M]\Omega_M,\;w_M[g_M]\Omega_M>\;\times\;
\sum_{m_1,m_2\in \mathbb{Z}_M}\; f_M(m_1) \; g_M(m_2)
\int\; dx\; F(x)\;
\nonumber\\ &&
[
\hat{\kappa}^{1\prime}_{s,M}(x;m_1,m_2)
+\hat{\kappa}^{2\prime}_{s,M}(x;m_1,m_2)
+\kappa^2_{s,M}(x;m_1,m_2)^\ast]
\ea
Comparing (\ref{5.29}) and (\ref{5.25}) we obtain exact match iff for 
$j=1,2$
\ba \label{5.30}
&& \hat{\kappa}^{j\prime}_{s,M}(x;m_1,m_2)=
\kappa^j_{s,M}(x;m_1,m_2)\;\;
\Leftrightarrow\;\;
\int\; dy\int\; dz \kappa^j_s(x;y,z)\; \sigma_M(y,m_1)\;\sigma_M(z,m_2)
\nonumber\\ 
&=&
\sum_{\hat{m}_1,\hat{m}_2}\;\hat{\kappa}^j_{s,M}(x;\hat{m}_1,\hat{m}_2)
\; 
\hat{\sigma}_M(\hat{m}_1,m_1)
\hat{\sigma}_M(\hat{m}_2,m_2)
\ea
which determines the discrete kernels 
$\hat{\kappa}^j_{s,M}(x;\hat{m}_1,\hat{m}_2)$ in terms of the continuum 
kernels $\kappa^j_s(x,;y,z)$. 

The question is, whether the flow $n\mapsto [D^{(n)}(F)]_M$ starting from 
(\ref{5.2}) actually yields this fixed point. Before we answer this question
we note that (\ref{5.27}) is simply not of the form
\be \label{5.31}
\int\; dx\; F(x)\; \sum_m\; E_M(x;m)\; D_{s,M}(m)
\ee 
which would yield a natural map (kernel) $E_M;\; L\mapsto L_M$, see the 
discussion of item 1., viewpoint i. in section \ref{s4}. 
It is not even of the form 
\be \label{5.32}
\int\; dx\; F(x)\; \sum_{m_1,m_2}\; E_M(x;m_1,m_2)\; D_{s,M}(m_1,m_2)
\ee 
in terms of a bi-kernel $E_M:\; L\mapsto L_M\times L_M$ because there 
are three independent monomials of annihilation and creation operators 
involved, not only one. Thus, blocking 
from the continuum does not give rise to such a {\it natural} 
kernel or bi-kernel 
which would allow us to consider the discretised constraints as $[C(F)]_M$
as smeared with a discretised function or bi-function. However, one 
may introduce such an interpretation {\it by hand} by restricting 
$F$ to be of the form $\hat{I}_M F_M$ where $\hat{I}_M$ should be sufficiently
differentiable and has all the properties of $I_M$, 
see again the discussion of item 1., viewpoint ii. in section \ref{s4}.  
Such $\hat{I}_M$ will be indeed be provided in \cite{27}.       
  
To study the actual flow of the constraints we note that 
\be \label{5.33}
\kappa^1_s(x;y,z)=\kappa_s(x,y)^\ast\;\kappa_s(x,z),\;\;
\kappa^2_s(x;y,z)=\kappa_s(x,y)\;\kappa_s(x,z),\;\;
\kappa_s(x,y)=[Q_s\sqrt{2\omega}](x,y)
\ee
while $\sigma=(2\omega)^{-1/2}Q$ and $\sigma_M=\sigma\circ I_M$ so that  
\ba \label{5.34}
\kappa^1_{s,M}(x;m_1,m_2)&=&\kappa_{s,M}(x,m_1)^\ast\;\kappa_{s,M}(x,m_2),\;\;
\kappa^2_{s,M}(x;m_1,m_2)=\kappa_{s,M}(x,m_1)\;\kappa_s(x,m_2),\;\;
\nonumber\\ 
\kappa_{s,M}(x,m)&=&[\kappa_s\circ \sigma_M](x,m)=[Q_s\circ I_M](x,m)
\ea
Accordingly we conclude that 
\ba \label{5.35}
&&\hat{\kappa}^1_{s,M}(x;m_1,m_2)=\hat{\kappa}_{s,M}(x,m_1)^\ast\;
\hat{\kappa}_{s,M}(x,m_2),\;\;
\nonumber\\ &&
\hat{\kappa}^2_{s,M}(x;m_1,m_2)=\hat{\kappa}_{s,M}(x,m_1)\;\hat{\kappa}_s(x,m_2),\;\;
\hat{\kappa}_{s,M}(x,m)=[Q_s I_M \sqrt{2\omega_M}](x,m)
\ea
because with $\hat{\sigma}_M=(2\omega_M)^{-1/2} Q_M$ we have 
\be \label{5.36}
[\hat{\kappa}_{s,M}\circ \hat{\sigma}_M](x,m)=[Q_s I_M Q_M](x,m)=
[Q_s Q I_M](x,m)=[Q_s I_M](x,m)=\kappa_{s,M}(x,m)
\ee
To see whether these fixed point values of the kernels are reached from 
the initial discretisation we write 
\ba \label{5.37}
&&\hat{\kappa}^{1(n)}_{s,M}(x;m_1,m_2)=\hat{\kappa}^{(n)}_{s,M}(x,m_1)^\ast\;
\hat{\kappa}^{(n)}_{s,M}(x,m_2),\;\;
\nonumber\\ &&
\hat{\kappa}^{2(n)}_{s,M}(x;m_1,m_2)=
\hat{\kappa}^{(n)}_{s,M}(x,m_1)\;\hat{\kappa}^{(n)}_s(x,m_2),\;\;
\hat{\kappa}^{(n)}_{s,M}(x,m)=[Q_s I_M \sqrt{2\omega^{(n)}_M}](x,m)
\ea
and by the literally identical calculation we obtain 
\be \label{5.38}
\hat{\sigma}_M^{(n)}(x,m)=[2\omega^{(n)}_M]^{-1/2} Q_M
\ee
in terms of which the flow equation reads
\ba \label{5.38} 
&& \sum_{\hat{m}_1,\hat{m}_2}\;\kappa^{j(n+1)}_{s,M}(x,\hat{m}_1,\hat{m}_2)\;
\hat{\sigma}^{(n+1)}_M(\hat{m}_1,m_1)\;
\hat{\sigma}^{(n+1)}_M(\hat{m}_2,m_2)\;
\nonumber\\
&=& \sum_{\hat{m}_1',\hat{m}_2'}
\kappa^{j(n)}_{s,M'}(x,\hat{m}'_1,\hat{m}'_2)\;
(\hat{\sigma}^{(n)}_{M'}\circ I_{MM'}(\hat{m}'_1,m_1)\;
(\hat{\sigma}^{(n)}_{M'}\circ I_{MM'}(\hat{m}'_2,m_2)\;
\ea
which is equivalent to 
\be \label{5.39}
\hat{\kappa}^{(n+1)}_{s,M}\circ \hat{\sigma}^{(n+1)}_M
=\hat{\kappa}^{(n)}_{s,M¨'}\circ \hat{\sigma}^{(n)}_{M'} \circ I_{MM'}
\ee
or 
\be \label{5.40}
\hat{\kappa}^{(n+1)}_{s,M}\circ Q_M
=\hat{\kappa}^{(n)}_{s,M¨'}\circ 
[\omega^{(n)}_{M'}]^{-1/2} \circ I_{MM'}
[\omega^{(n)}_M]^{1/2}\circ Q_M 
\ee
where the sequence $n\mapsto \omega^{(n)}_M$ was constructed explicitly
from the measure flow and satisfies for $M'(M)=2M$
\be \label{5.42}
I_{MM'(M)}^\dagger\; [\omega^{(n)}_{M'(M)}]^{-1}\;
I_{MM'(M)}= [\omega^{(n+1)}_M]^{-1}\;
\ee
Starting with 
\be \label{5.43}
\hat{\kappa}^{(0)}_{s,M}=I_M \; Q^{(0)}_{s,M} [\omega^{(0)}_M]^{1/2}
\ee
one finds from (\ref{5.40}) using the consistency of the maps 
$I_{M_2 M_3} I_{M_1 M_2}=I_{M_1 M_3}$ for $M_1<M_2<M_3$
\be \label{5.44}
\hat{\kappa}^{(n)}_{s,M}=I_{2^n M} \; Q^{(0)}_{s,2^n M}
I_{M, 2^n M} [\omega^{(n)}_M]^{1/2}
\ee
Taking the limit $n\to \infty$ we get, due to limit values
$I_\infty=1_L,\; 
Q^{(0)}_{s,\infty}=Q_s, \; I_{M,\infty}=I_M,\; \omega^{(\infty)}_M$ formally
\be \label{5.45} 
\hat{\kappa}^{(\infty)}_{s,M}=
\hat{\kappa}_{s,M}
\ee
However, it must be shown if and in what sense the sequence 
(\ref{5.44}) actually 
runs into the limit (\ref{5.45}) which coincides with that blocked 
from the continuum. This will be done in the next section.

\section{Discrete Virasoro Algebra}
\label{s6}

The current section is the most important one of the present paper as 
it answers the question whether the continuum algebra is visible at finite 
resolution, how large its finite resolution anomaly is and in what sense 
that anomaly is simply a finite resolution artefact and converges to zero
as we increase the resolution.\\
\\
We thus consider the finite resolution $M$ constraint operators on 
${\cal H}_M$
\be \label{6.1} 
D_{sM}(F):=J_M^\dagger\;D_s(F)\; J_M
\ee
and compute the finite resolution anomaly
\be \label{6.2}
\alpha_M(F,s;G,t):=
[D_{sM}(F),D_{tM}(G)]-J_M^\dagger\;[D_s(F),D_t(G)]\;J_M 
=-J_M^\dagger[ D_s(F)\; P_M^\perp\; D_t(G)
-D_t(G)\; P_M^\perp\; D_s(F)] \; J_M
\ee
where 
\be \label{6.3}
P_M^\perp=1_{{\cal H}}-P_M,\;\; P_M=J_M\; J_M^\perp=P_M^2=P_M^\dagger
\ee
is an orthogonal projection thanks to the isometry 
$J_M^\dagger J_M=1_{{\cal H}_M}$. The finite resolution anomaly vanishes 
only when the constraint operators preserve the subspaces $P_M\;{\cal H}$
of ${\cal H}$ which is generically not the case and certainly for PFT it is 
not.

Heuristically the anonaly vanishes as we increase the resolution 
$M\to\infty$ as we expect that $P_M^\perp \to 0$. The rest of this section 
is devoted to showing that this is the case rigorously in a suitable 
operator topology. In fact showing that $\alpha_M(s,F;t,G)$ as $M\to \infty$
is a delicate issue and must be defined appropriately. This
is because 
we change the Hilbert space ${\cal H}_M$ on which $\alpha_M$ is defined. 
Hence we cannot simply probe the anomaly, say with respect to the 
weak operator topology on ${\cal H}_M$, that is, 
fixing $\psi_M,\psi'_M\in {\cal H}_M$, considering 
the matrix elements 
\be \label{6.4}
<\psi_M,\;\alpha_M(s,F;t,G)\psi'_M>_{{\cal H}_M} 
\ee
and taking $M\to \infty$ at fixed $\psi_M,\psi'_M$ as these depend themselves
on $M$. However, what we can do is to consider fixed 
$\psi,\psi'\in {\cal H}$ independent of $M$ and probe the anomaly with 
$\psi_M:=J_M^\dagger\psi,\; 
\psi'_M:=J_M^\dagger\psi'$. Accordingly we study the large $M$ behaviour 
of   
\be \label{6.5}
<J_M^\dagger\; \psi,\;\alpha_M(s,F;t,G)\; J_M^\dagger\psi'>_{{\cal H}_M} 
\ee
It will be sufficient to study one of the two terms in (\ref{6.2}) i.e. the 
matrix element 
\be \label{6.6}
<\psi,\; P_M \; D_s(F)\; P_M^\perp\; D_t(G)\; P_M\; \psi'>_{{\cal H}}
=<D_s(F)\; P_M \psi,\; P_M^\perp\; D_t(G)\; P_M\; \psi'>_{{\cal H}}
\ee
where used the symmetry of all operators involved.

There are several issues with (\ref{6.6}) that require clarification:
First of all, one would like to take $\psi,\;\psi'$ from the dense domain 
$\cal D$ given by the span of the Weyl vectors $w[f]\Omega$, however, 
to be useful we need an explicit formula for 
$J_M^\dagger\psi,\; P_M\psi$ for $\psi\in {\cal D}$ which is not available
from \cite{4,6,7}. We derive this formula below. Next, as expected, the range 
of $J_M^\dagger {\cal D}$ is in ${\cal D}_M$ which is the span of the 
$w[I_M\; f_M]\Omega$ which is dense in $P_M {\cal H}$. However, as 
$I_M f_M$ is a step function, it is not clear that 
$D_s(F) w[I_M f_M]\Omega$ is well defined, i.e. a normalisable element of
$\cal H$. It is for this reason that we considered also the case of 
discontinuous functions $f$ such as $I_M f_M$ as the domain of the constraint
operators in section \ref{s2} and we showed that after suitable 
regularisation we have indeed $D_s(F)\; w[I_M f_M]\Omega \in {\cal H}$. 
Finally, the image of ${\cal D}$ or $P_M {\cal D}$ is not invariant under 
the constraints so that evaluation of the matrix elements of $P_M^\perp$ 
between vectors in $\cal D$ is 
again not directly possible. In fact, in order to evaluate $P_M^\perp$ 
on say $D_s(F)\; P_M\; w[f]\Omega$ one would need to know how to write it as
a linear combination of the $w[g]\Omega$, a task which has no 
obvious solution. One could think that one can avoid this 
complication and use the fact that $\cal D$ is dense in $\cal H$.
Thus given $\epsilon$ we find $\tilde{\psi}\in {\cal D}$ which differs 
in norm from $D_s(F)\; P_M w[f]\Omega$ by at most $\epsilon$. If that 
$\tilde{\psi}$ would only depend on $s,F,\epsilon$ one could indeed 
restrict consideration to the limit of the matrix elements 
$<\tilde{\psi},P_M^\perp\; \tilde{\psi}'>$ with 
$\tilde{\psi},\tilde{\psi}'\in {\cal D}$ because $||P_M^\perp||=1$ 
is bounded. 
Unfortunately, such $\tilde{\psi}$ does depend on $M$ and without explicitly 
knowing how it does so, it is not possible to estimate the limit of 
(\ref{6.6}). The fact that also $||P_M||=1$ does not help as $P_M$ 
stands between $D_s(F)$ and $\psi$.

We are therefore forced to have a detailed look at (\ref{6.6}). A 
simplification can be obtained by observing that 
\ba \label{6.7}
&&|<D_s(F)\; P_M \psi,\; P_M^\perp\; D_t(G)\; P_M\; \psi'>|
\le <D_s(F)\; P_M \psi,\; P_M^\perp\; D_s(F)\; P_M\; \psi>^{1/2}
\;
\nonumber\\ &&
\le <D_t(G)\; P_M \psi',\; P_M^\perp\; D_t(G)\; P_M\; \psi'>^{1/2}
\ea
thanks to the CS inequality and the projector property 
$(P_M^\perp)^2=(P_M^\perp)^\dagger=P_M^\perp$. Thus (\ref{6.6}) converges 
to zero as $M\to \infty$ for all $s,F,\psi,t,G,\psi'$ if and only if
\be \label{6.8}
<D_s(F)\; P_M \;\psi,\; P_M^\perp\; D_s(F)\; P_M\; \psi>
\ee
converges to zero for all $s,F,\psi\in {\cal D}$: That 
convergence of (\ref{6.6}) implies convergence of (\ref{6.8}) follows 
by choosing $t=s,G=F,\psi'=\psi$. Next convergence of (\ref{6.8}) for 
all $\psi\in {\cal D}$ implies in particular convergence of 
\be \label{6.9}
<D_s(F)\; P_M \;w[f]\Omega,\; P_M^\perp\; D_s(F)\; P_M\; w[f]\Omega>
\ee
for the choice $\psi=w[f]\Omega$ and conversely convergence of (\ref{6.9})
implies convergence of (\ref{6.8}) for finite linear combinations of 
the $w[f]\Omega$, that is, general $\psi\in {\cal D}$ again by the CS 
inequality.\\
\\
Accordingly we will prove that (\ref{6.9}) converges to zero. Our first 
task is to compute $P_M w[f]\Omega$. We begin by computing 
$J_M^\dagger \; w[f]\Omega$ 
\ba \label{6.10}
&& <w_M[g_M]\Omega_M,\; J_M^\dagger\; w[f]\Omega>_{{\cal H}_M}
<J_M \;w_M[g_M]\Omega_M,\;w[f]\Omega>_{{\cal H}}
\nonumber\\  
&=&  
<w[I_M g_M]\Omega,\;w[f]\Omega>_{{\cal H}}
=<\Omega,\;w[f-I_M g_M]\Omega>_{{\cal H}}
= \exp(-\frac{1}{2}\; C(f-I_M g_M,f-I_M g_M))
\ea
where we have written out the continuum covariance 
\be \label{6.11}
2\; C=
Q^\perp\; \omega_0^{-1} \; Q^\perp+
Q\; \omega^{-1} \; Q
\ee
as a symmetric bilinear form on $L\times L$. We can also consider it as an 
operator defined by 
\be \label{6.12}
<f,\;C\; g>_L:=C(f,g)
\ee
We will make use of these two meanings of $C$ as appropriate, it is clear from
the context which meaning is used respectively. We also remind of the 
covariance at resolution $M$ 
\be \label{6.13}
2\;C_M=I_M^\dagger\; 2\;C\; I_M=  
Q_M^\perp\; \omega_0^{-1} \; Q_M^\perp+
Q_M\; \omega_M^{-1} \; Q_M
\ee
where equivariance $Q \; I_M=I_M\; Q_M$ was used. Note that both $C, C_M$ 
considered as operators on $L, L_M$ respectively have, in contrast to 
$\omega,\omega_M$ an inverse, explicitly
\be \label{6.13}
\frac{1}{2} \;C_M^{-1}=I_M^\dagger\; 2\;C\; I_M=  
Q_M^\perp\; \omega_0 \; Q_M^\perp+
Q_M\; \omega_M \; Q_M
\ee
and similar for $C^{-1}$. 

We make the Ansatz 
\be 
\label{6.14}
J_M^\dagger \; w[f]\Omega=\kappa_M(f)\; w_M[f_M(f)]\;\Omega_M
\ee
for numbers $\kappa_M(f)$ and vectors $f_M(f)\in L_M$ to be determined. 
Plugging (\ref{6.14}) into (\ref{6.10}) we find 
\be \label{6.15}
\exp(-\frac{1}{2}\; C(f-I_M g_M,f-I_M g_M))
=\kappa_M(f)\;
\exp(-\frac{1}{2}\; C_M(f_M(f)-g_M,f_M(f)-I_M g_M))
\ee
which is uniquely solved by 
\be \label{6.16}
f_M(f)=C_M^{-1}\; I_M^\dagger\; C\; f,\;
\kappa_M(f)=\exp(-\frac{1}{2}[C(f,f)-C_M(f_M(f),f_M(f))])
\ee
Note that $\kappa_M(f)$ can be simplified
\ba \label{6.17}
&& C_M(f_M(f),f_M(f))=C(I_M\; f_M(f),I_M\; f_M(f))
=<I_M\; f_M(f), C\; I_M \; f_M(f)>_L
\nonumber\\
&=& <f_M(f), I_M^\dagger\; C\; I_M \; f_M(f)>_{L_M}
= <f_M(f), C_M\; f_M(f)>_{L_M}
\nonumber\\
&=& <f_M(f), I_M^\dagger\; C\;f>_{L_M}
=<f,\;[C\; (I_M \; C_M^{-1}\; I_M^\dagger)\; C]\;f>_L
\ea
It follows 
\be \label{6.18}
P_M \; w[f]\Omega = \kappa_M(f)\; w[f^M(f)]\;\Omega,\;\;
f^M(f)=I_M \; f_M(f)=(I_M \; C_M^{-1}\; I_M^\dagger)\; C\;f     
\ee
It is instructive to verify the projection property $P_M^2=P_M$ and 
the isometry property $J_M^\dagger\; J_M=1_{{\cal H}_M}$ which relies 
on $\kappa_M(I_M f_M)=1$ and $f^M(I_M f_M)=I_M f_M$ for any $f_M\in L_M$. 
    
The next task is to compute $D_s(F) \; P_M\; w[f]\Omega$ which given 
(\ref{6.18}) can be done of course uising the explicit expression of 
$D_s(F)$ in terms of creation and annihilation operators. However, to be 
useful, we must write $D_s(F)\; P_M\; w[f]\Omega$ in the form of 
linear combinations of $w[h]\Omega$ again because in order to apply 
$P_M^\perp$ to it, whose action follows from (\ref{6.18}), its action 
is only known 
in closed form on vectors in $\cal D$ and not on Fock states. 
The other option would be to expand $P_M\; w[f]\Omega$ into 
of Fock states. While this is possible, it leads to very complex expressions.
We therefore choose the former route which also has the advantage to 
maximally benefit from the identity $P_M^\perp \; P_M=0$.  

We note that (we pick the C sector for definiteness and focus only on the 
correpsonding contribution to the constraints)
\ba \label{6.19}
&& w[h]\Omega
=\exp(i<h,\Phi>)\;\Omega=
\exp(i<C^{-1/2}\;h,\;A_C^\dagger+A_C>_L)\;\Omega
\nonumber\\
&=& \exp(-\frac{1}{2}<h,\; C\;h>\; 
\exp(i<C^{1/2}\;h,\;A_C^\dagger+>_L)\;\Omega
\ea
using well known Fock space techniques (BCH formula). Here we have denoted
the annihilation operator of the $C$ sector by $A_C$ in order not to confuse 
it with the covariance $C$. Thus we find the functional derivatives
\ba \label{6.20}
&&\frac{\delta}{\delta h(y)}\; w[C^{-1}h+g]\Omega
=[-g(y)-(C^{-1}h)(y)+i(C^{-1/2}A_C^\dagger)(y)]\; w[C^{-1}h+g]\Omega
\nonumber\\
&&\frac{\delta^2}{[\delta h(y)][\delta h(z)]}\; w[C^{-1}h+g]\Omega
=\{-C^{-1}(y,z)+
[-g(y)-(C^{-1}h)(y)+i(C^{-1/2}A_C^\dagger)(y)]
\nonumber\\ &&
\;[-g(z)-(C^{-1}h)(z)+i(C^{-1/2}A_C^\dagger)(z)]
\; w[C^{-1}h+g]\Omega
\ea
i.e. at $h=0$          
\ba \label{6.20}
&&(\frac{\delta}{\delta h(y)}\; w[C^{-1}h+g]\Omega)_{h=0}
=[-g(y)+i(C^{-1/2}A_C^\dagger)(y)]\; w[g]\Omega
\\
&&(\frac{\delta^2}{[\delta h(y)][\delta h(z)]}\; w[C^{-1}h+g]\Omega)_{h=0}
=\{-C^{-1}(y,z)+
[-g(y)+i(C^{-1/2}A_C^\dagger)(y)]
\;[-g(z)+i(C^{-1/2}A_C^\dagger)(z)]
\;\}\; w[g]\Omega
\nonumber
\ea 
Here we used that all expressions just depend on creation operators 
which mutually commute. 

Recall the constraint operator 
\ba \label{6.21}
&& -D_s(F)=\int\; dx\; F(x)\;\int\; dy\; \int\; dz\;\{
Q_s(x,y)\;Q_s(x,z) \; (C^{-1/2}A_C)(y)\; (C^{-1/2}A_C)(z)\; 
\\
&& +Q^\ast_s(x,y)\;Q^\ast_s(x,z) \; (C^{-1/2}A_C^\dagger)(y)\; 
(C^{-1/2}A_C^\dagger)(z)\; 
-2\; Q^\ast_s(x,y)\;Q_s(x,z) \; (C^{-1/2}A_C^\dagger)(y)\; (C^{-1/2}A_C)(z)\; 
\}
\nonumber
\ea
where $Q_s(x,y)$ is the integral kernel of the projection $Q_s$. We have 
explicitly 
\ba \label{6.22}
&& [C^{-1/2}A_C](y) \; w[g]\Omega
=w[g]\;(w[-g]\;[C^{-1/2}A_C](y) \; w[g])\Omega
=w[g]\;([C^{-1/2}A_C](y)-i[\phi[g],(C^{-1/2} A_C)(y)])\Omega
\nonumber\\
&=& -i \; w[g]\;[A_C^\ast[C^{1/2}g],(C^{-1/2} A_C)(y)])\Omega
=-i \; g(y)\; w[g]\;\Omega
\ea
whence 
\ba \label{6.23}
&& -D_s(F)\; w[g]\Omega
=\int\; dx\; F(x)\;\int\; dy\; \int\; dz\;\{
-Q_s(x,y)\;Q_s(x,z) \; g(y)\; g(z)\; 
\nonumber\\ &&
+Q^\ast_s(x,y)\;Q^\ast_s(x,z) \; (C^{-1/2}A_C^\dagger)(y)\; 
(C^{-1/2}A_C^\dagger)(z)\; 
-2i\; Q^\ast_s(x,y)\;Q_s(x,z) \; (C^{-1/2}A_C^\dagger)(y)\; g(z)\; 
\}\;
w[g]\Omega
\nonumber\\
&=& \int\; dx\;\int\; dy\; \int\; dz\;\{
\{(-Q_s(x,y)\;Q_s(x,z)+Q^\ast_s(x,y)\;Q^\ast_s(x,z)
-2\;Q_s(y,z)^\ast Q(x,z)) \;g(y)\; g(z)
-C^{-1}(x,y)\}
\nonumber\\
&& +Q^\ast_s(x,y)\;Q^\ast_s(x,z) \;
\{(C^{-1/2}A_C^\dagger)(y)\; (C^{-1/2}A_C^\dagger)(z)\;
+2i\; g(y)\;(C^{-1/2}A_C^\dagger)(z)+C^{-1/2}(y,z)-g(y)\;g(z)\}
\nonumber\\
&& -2\;\{ (Q^\ast_s(x,y)\;Q_s(x,z)+Q^\ast_s(x,y)\;Q^\ast_s(x,z)) \;
\;[i(C^{-1/2}A_C^\dagger)(y)-g(y)]\; g(z)\; \}\;
\}\;\;
w[g]\Omega
\ea
with $Q=Q_s+Q_s^\ast$.
We evaluate (\ref{6.23}) for $g=f^M(f)$, multiply it from the left with 
$\kappa_M(f)\; P_M^\perp$ and use (\ref{6.20}) to obtain the identity
\ba \label{6.24}
&& P_M^\perp\; D_s[F]\; P_M\; w[f]\Omega
= \{P_M^\perp\;\int\; dx\; F(x)\;\int\; dy\; \int\; dz
\\
&& \{Q_s(x,y)^\ast Q_s(x,z)^\ast\; \frac{\delta^2}{[\delta h(y)]\;[\delta h(z)]}
-2\;Q_s(x,y)^\ast Q(x,z)\; [f^M(f)](z)\;
\frac{\delta}{\delta h(y)}\;\}
\kappa_M(f)\; w[C^{-1}\; h+f_M(f)]\Omega\}_{h=0}
\nonumber
\ea
where the terms in (\ref{6.23}) that do not involve creation operators could 
be dropped because at $h=0$ we get 
$P_M^\perp\;\kappa_M(f)w[C^{-1}h+f_M(f)]\Omega=P_M^\perp P_M w[f]\Omega=0$.

Formula (\ref{6.24}) is the desired expression because $P_M^\perp$ can be 
pulled past the functional derivatives where it hits 
$w[C^{-1} h+f^M(f)]\Omega$ and can be  evaluated. 
Let $h'=C^{-1} h,\;g=f^M(f)$. Then due to the projector property 
$f^M(g)=g$ and $\kappa_M(g)=1$ whence 
\ba \label{6.25}
P_M^\perp\; w[h'+g]\Omega
&=& w[h'+g]\Omega-\kappa_M(h'+g)\; w[f^M(h'+g)]\Omega
\nonumber\\
f^M(h'+g) &=& f^M(h')+g     
\nonumber\\
\kappa_M(h'+g) &=& \kappa_M(h')\; \kappa_M(g)\; 
\exp(<h',C(1-I_M C_M^{-1} I_M^\dagger)\; g>_L
=\kappa_M(h')
\ea
therefore 
\be \label{6.26}
\kappa_M(f)\; P_M^\perp\; w[h'+g]\;\Omega
=\kappa_M(f)\; (w[h'+g]-\kappa_M(h')\; w[f^M(h')+g]\;\Omega
=[w[h']-\kappa_M(h')]\; w[f^M(h')]\;P_M\;w[f]\;\Omega
\ee
We can now evaluate (\ref{6.8})
\ba \label{6.27}
&&||P_M^\perp\; D_s(F)\; P_M\; w[f]\Omega||^2
=\{\;\;\;\;
\int\;dx\; F(x)\;\int\; dy\;\int\; dz\;
\int\;dx'\; F(x')\;\int\; dy'\;\int\; dz'\;
\\
&&
\{Q_s(x,y)\;Q_s(x,z)\frac{\delta^2}{[\delta h(y)]\;[\delta h(z)]}
-2\;Q_s(x,y)\;Q(x,z)\;g(z)\;\frac{\delta}{\delta h(y)}\}
\nonumber\\
&&
\{Q^\ast_s(x',y')\;Q^\ast_s(x',z')\frac{\delta^2}{[\delta \hat{h}(y')]\;
[\delta \hat{h}(z')]}
-2\;Q^\ast_s(x',y')\;Q(x',z')\;g(z')\;\frac{\delta}{\delta \hat{h}(y')}\}
\nonumber\\
&& <[w[h']-\kappa_M(h')]\; w[f^M(h')]\;P_M\;w[f]\;\Omega,\;
[w[\hat{h}']-\kappa_M(\hat{h}')]\; w[f^M(\hat{h}')]\;P_M\;w[f]\;
\Omega>_{{\cal H}}
\;\;\;\;\}_{h=\hat{h}=0}
\nonumber
\ea
with $g=f^M(f),\; h'=C^{-1} h,\; \hat{h}'=C^{-1} \hat{h}$. We have 
\ba \label{6.28} 
&& <[w[h']-\kappa_M(h')]\; w[f^M(h')]\;P_M\;w[f]\;\Omega,\;
[w[\hat{h}']-\kappa_M(\hat{h}')]\; w[f^M(\hat{h}')]\;P_M\;w[f]\;\Omega>
\\
&=&
\kappa_M(f)^2\; 
 <[w[h']-\kappa_M(h')]\; w[f^M(h')]\;w[g]\;\Omega,\;
[w[\hat{h}']-\kappa_M(\hat{h}')]\; w[f^M(\hat{h}')]\;w[g]\;\Omega>
\nonumber\\
&=&
\kappa_M(f)^2\; 
 <[w[h']-\kappa_M(h')]\; w[f^M(h')]\;\Omega,\;
[w[\hat{h}']-\kappa_M(\hat{h}')]\; w[f^M(\hat{h}')]\;\Omega>
\nonumber\\
&=&
\kappa_M(f)^2\; 
 <P_M^\perp \; w[h']\;\Omega,\;P_M^\perp\;w[\hat{h}']\;\Omega>
\nonumber\\
&=&
\kappa_M(f)^2\; 
 <P_M^\perp \; w[h']\;\Omega,\;P_M^\perp\;w[\hat{h}']\;\Omega>
\nonumber\\
&=&
\kappa_M(f)^2\; 
<w[h']\;\Omega,\;P_M^\perp\;w[\hat{h}']\;\Omega>
\nonumber\\
&=&
\kappa_M(f)^2\; 
\; 
<\Omega,\;[w[\hat{h}'-h']-\kappa_M(\hat{h}')\;
w[f^M(\hat{h}')-h']\;\Omega>
\nonumber\\
&=&
\kappa_M(f)^2\; 
\;[\exp(-\frac{1}{2}\;<\hat{h}'-h',C(\hat{h'}-h')>)
-\kappa_M(\hat{h}')\;
\;\exp(-\frac{1}{2}\;<f^M(\hat{h}')-h',C(f^M(\hat{h'})-h')>)]
\nonumber
\ea
Before evaluating the functional derivatives we can simplify (\ref{6.28})
\ba \label{6.29}
&& <h',C\; f^M(\hat{h}')>=<f^M(h'),C\;\hat{h}'>
\nonumber\\
&& <f^M(\hat{h}'),C\; f^M(\hat{h}')>
=<\hat{h}',C\; f^M(\hat{h}')>
\nonumber\\
&&
\kappa_M(\hat{h}')\;
\;\exp(-\frac{1}{2}\;<f^M(\hat{h}')-h',C(f^M(\hat{h'})-h')>)]
\nonumber\\
&=&
\exp(-\frac{1}{2}[
<\hat{h}',C[\hat{h}'-f^M(\hat{h}')]>
+<f^M(\hat{h}')-h',\;C\;[f^M(\hat{h}')-h']>])
\nonumber\\
&=&
\exp(-\frac{1}{2}[<[\hat{h}'-h']\;C[\hat{h}'-h']+
2\;<h',C(\hat{h}'-f^M(\hat{h}')>])
\ea
Accordingly, (\ref{6.28}) can be rewritten as (reintroducing $h=C\;h',\;
\hat{h}=C\; \hat{h}'$)
\be \label{6.30}
\kappa_M(f)^2\; 
\exp(-\frac{1}{2}\;<\hat{h}-h,\;C^{-1}\;[\hat{h}-h]>)\;
[1-\exp(-<h,\;[C^{-1}-I_M\;C_M^{-1}\;I_M^\dagger]\;\hat{h}>)]
\ee
It will be convenient to define the symmetric kernels $K=C^{-1},\;\Delta K=
C^{-1}-I_M\; C_M^{-1}\; I_M^\dagger$. In carrying out the double, triple
and four-fold
functional derivative of (\ref{6.30}) at $h=\hat{h}=0$ we use arguments 
familiar from Wick's theorem in perturbative QFT: as (\ref{6.30}) is 
a linear combination of two exponentials $E(H)=\exp(B(H,H)/2)$ 
of a quadratic polynomial 
$B$ in $H=(h,\hat{h}')$, 
their derivatives are schematically
\ba \label{6.31}
&& E'=(BH)\; E,\;
E^{\prime\prime}=[B+(BH)^2]\; E,\;
E^{\prime\prime\prime}=[3\;B^2 H+(BH)^3]\; E,\;
\nonumber\\
&& E^{\prime\prime\prime\prime}=
[3\;B^2 +3(BH)^2\; B+3\;(B^2 H)(BH)+(BH)^4]\; E,\;
\ea
so that at $H=0$ only second and fourth derivatives survive. To simplify 
the notation we set 
\be \label{6.33}
E_1:=\exp(-\frac{1}{2}\;<\hat{h}-h,\;K\;[\hat{h}-h]>),\;
E_2:=\exp(-<h,\;[\Delta K]\;\hat{h}>),\;
E_{j,y}:=\frac{\delta}{\delta h(y)},\;
E_{j,y'}:=\frac{\delta}{\delta \hat{h}(y')},\;
\ee
with $j=1,2$ and similar for $z,z'$. Then 
\ba \label{6.34}
&& (E_1 E_2)_{,y y'}=E_{1,yy'}\; E_2+E_1\; E_{2,yy'}
+E_{1,y}\; E_{2,y'}+E_{1,y'}\; E_{2,y}
\nonumber\\
&& (E_1 E_2)_{,y y' z z'}=
[E_{1,yy'zz'}\; E_2+E_{1,yy'}\; E_{2,zz'}]
+[E_{1,zz'}\; E_{2,yy'} +E_1\; E_{2,yy'zz'}]
\nonumber\\ &&
+[E_{1,yz}\; E_{2,y'z'}=E_{1,yz'}\; E_{2,y'z}]
+[E_{1,y'z}\; E_{2,yz'}+E_{1,y'z'}\; E_{2,yz}]
+...
\ea
where $...$ denotes odd order derivatives which vanish at $H=0$. We have at 
$H=0$
\ba \label{6.35}
&& E_{1,yz}=-K(y,z),\;E_{1,y'z'}=-K(y',z'),\;E_{1,yz'}=K(y,z'),\;
\nonumber\\
&&
E_{2,yz}=0,\;E_{2,y'z'}=0,\;E_{2,yz'}=[\Delta K](y,z'),\;
\nonumber\\
&&
E_{1,yy'zz'}=
K(y,z)\;K(y',z')+K(y,y')\;K(z,z')+
K(y,z')\;K(z,y')
\nonumber\\
&&
E_{2,yy'zz'}=
[\Delta K](y,y')\;[\Delta K](z,z')
+[\Delta K](y,z')\;[\Delta K](z,y')
\ea
Collecting all terms we find at $H=0$
\ba \label{6.3}
&& [E_1(1-E_2)]_{,yy'}=[\Delta K](y,y') 
\\
&&
[E_1(1-E_2)]_{,yy'zz'}=
K(y,y')\;[\Delta K](z,z')
+K(y,z')\;[\Delta K](z,y')
K(z,y')\;[\Delta K](y,z')
\nonumber\\ &&
-[\Delta K](y,y'\;[\Delta K](z,z')\;
-[\Delta K](y,z'\;[\Delta K](z,y')
\ea
where importantly both terms proportional to $E_{1,yy'zz'}$ have cancelled 
so that all functional derivatives contain at least one factor of
$\Delta K$ which we expect to imply the convergence to zero of 
(\ref{6.8}) which now can be vastly simplified to 
\be \label{6.37}
\kappa_M(f)^2\;\int\; dx\; F(x)\;\int\; dx'\; F(x')\{
\{3\;K_s(x,x')\;[\Delta K]_s(x,x')-2\;([\Delta K]_s(x,x'))^2 
+4 \; g'(x) g'(x')\; [\Delta K]_s(x,x')\}
\ee
where using $Q^\ast_s(y,z)=Q_{-s}(y,z)=Q_s(z,y)$
\be \label{6.38}
K_s(x,x')=
\int\; dy\;\int\; dz\; Q_s(x,y)\; Q_s(x',z) \; K(y,z)
=[Q_s\; K\; Q_s](x,x')
\ee
and similar for $[\Delta K]_s(x,x')$. Here  
\be \label{6.39}
g'(x)=[Q\; I_M\; C_M^{-1}\; I_M^\dagger\; C\; f](x)=(Q(K-[\Delta K])C f)(x),
\; \kappa_M(x)=\exp(-\frac{1}{2}<Cf,[\Delta K] Cf>)
\ee
Since $P_M$ is a projection we have $||P_M||=1$ thus 
\be \label{6.40}
||P_M\; w[f]\Omega||=\kappa_M(f)\;||w[f^M(f)]\;\Omega||=\kappa_M(f)||
\le ||P_M||\;||w[f]\Omega||=1
\ee
and it will be sufficient to show that the integral term in (\ref{6.37}) 
converges. Also we focus on $s=+$ the case $s=-$ being completely analogous. 
Obviously then, the convergence or not of (\ref{6.8}) 
rests on the properties of 
$\Delta_K$ and $g'$. We begin with the term 
\be \label{6.41}
\int\;dx\;\int \; dx'\; F(x)\;F(x')\;
K_+(x,x')\;[\Delta K]_+(x,x')
=\int\;dx\;\int \; dx'\; F(x)\;F(x')\;
K_+(x,x')\;[\Delta K]_-(x',x)
\ee
where in the second step we used that $[\Delta K](y,z)=[\Delta K](z,y)$ and 
$Q_s(y,z)=Q^\ast_s(z,y)=Q_{-s}(z,y)$. We expand into the Fourier basis
\ba \label{6.42}
K_+(x,x')
&=&\sum_{n,n'\in \mathbb{Z}}\; e_n(x)\;<e_n,\; Q_+\; K\;Q_+\;e_{n'}>\;
e_{-n'}(x')
=\sum_{n,n'>0}\; e_n(x)\;<e_n, K\;\;e_{n'}>\;e_{-n'}(x')
\nonumber\\ 
{[}\Delta K]_-(x',x) &=&
\sum_{n,n'<0}\; e_n(x')\;<e_n, [\Delta K]\;\;e_{n'}>\;e_{-n'}(x)
\nonumber\\
F(x) &=& \sum_{|n| <n_0} \; \hat{F}(n)\; e_n(x)=F^\ast(x),\;\hat{F}^\ast(n)=
\hat{F}(-n)
\ea
where we assume that {\it $F$ has compact momentum support} $|n|<n_0$. 
Presumably 
what follows can also be shown under milder decay assumptions on the 
Fourier modes $\hat{F}(n)$ (e.g. rapid decrease in $n\in \mathbb{Z}$) but 
we will be satisfied if convergence can be proved for this class of 
smearing functions of the constraint. Then (\ref{6.41}) turns into
\be \label{6.43}
\sum_{|n_1|,|n_2|<n_0} \hat{F}(n_1);\hat{F}^\ast(n_2)
\sum_{m,n>0} \; <e_m,K\; e_n>\; \sum_{m',n'<0}
[\Delta K]_{m',n'}\; \delta_{n_1+m-n'}\; \delta_{-n_2-n+m'}
\ee
This implies the constraints on the range of $m,n,m',n'$ 
\ba \label{6.44}
&& n'=n_1+m<0,\; m'=n_2+n<0,\; m=n'-n_1>0,\;n=m'-n_2>0\;\;
\nonumber\\ &&
\Rightarrow\;\;
0<m<-n_1<n_0,\;0<n<-n_2<n_0,\;0>n'>n_1>-n_0,\;0>m'>n_2>-n_0
\ea
thus the compact momentum support propagates to the $m,n,m',n'$ modes. For 
bounded values of $m,n$ the modulus of the 
matrix element $|<e_m, \; K\; e_n>|$ is uniformly bounded and 
we are left to study the behaviour of $<e_n,[\Delta K]\;e_{n'}>$ at 
fixed values of $n,n'\not=0$ (of equal sign). We have 
\ba \label{6.45}
&& <e_n, [\Delta K]\; e_{n'}>   
=<e_n, [C^{-1}-I_M\; C_M^{-1} I_M^\dagger\; e_{n'}>   
\nonumber\\
&=& 
2[\omega(n)\;\delta_{n,n'}-\sum_{\tilde{n}\in \mathbb{Z}_M}\;
\omega_M(\tilde{n})\;
<e_n,I_M\; e^M_{\tilde{n}}>\; <I_M e^M_{\tilde{n}},e_{n'}>
\ea
where in the second step we expanded into the spectral basis 
$e^M_{\tilde{n}}\in L_M$ of $\omega_M$ given by 
$e^M_{\tilde{n}}(m)=e_{\tilde{n}}(x^M_m),\;x^M_m=\frac{m}{M},\;
m\in \mathbb{Z}_M$. The eigenvalues $\omega_M(\tilde{n})$ follow from 
the definition $C_M=I_M^\dagger \; C\; I_M$ i.e. 
\be \label{6.46}
Q_M \omega_M^{-1} Q_M=I_M^\dagger\; Q\; \omega^{-1}\; Q\; I_M
\ee
from which 
\ba \label{6.47}
&& Q_M \omega_M^{-1} Q_M\;e^M_{\tilde{n}}
=\sum_{0\not=n,n'\in \mathbb{Z}}\;I_M^\dagger e_n \; 
<e_n,\omega^{-1}\;e_{n'}> \; <e_{n'},I_M e^M_{\tilde{n}}>
\nonumber\\
&=& \sum_{\tilde{n}'\in \mathbb{Z}_M}\; e^M_{\tilde{n}'}\;
\sum_{n\not=0}\; <I_M e^M_{\tilde{n}'}, e_n>\;\;\omega^{-1}(n)\; 
<e_n,I_M e^M_{\tilde{n}}>
\ea
Here we need the Fourier modes of the characteristic functions $\chi^M_m$ 
of the interval $[x^M_m,x^M_{m+1})$
\be \label{6.48}
(I_M^\dagger e_n)(m)=M<\chi^M_m,\;e_n>=M\; e_n(x^M_m)
\frac{e^{i\;k_M\;n}-1}{2\pi i\; n},\;k_M=\frac{2\pi}{M}
\ee
We note that (\ref{6.48}) does not have compact momentum support and also
does not decay rapidly. This has some bearing further below. It follows
\ba \label{6.49}
&& <e^M_{\tilde{n}},\;I_M^\dagger e_n>=
<I_M e^M_{\tilde{n}},\;e_n>=
\nonumber\\
&=& \sum_{m\in \mathbb{Z}_M}\; [e^M_{\tilde{n}}(m)]^\ast\; e^M_n(m)
\frac{e^{ik_M n}-1}{2\pi in}
=[\sum_m\; e^M_{n-\tilde{n}}(m)]\;\frac{e^{ik_M n}-1}{2\pi in}
\nonumber\\
&=& M\delta_{\tilde{n},\hat{n}}\;\frac{e^{ik_M n}-1}{2\pi in}
\ea
where $\hat{n}\in \mathbb{Z}_M$ and $n=\hat{n}+l M,\; l\in \mathbb{Z}$ 
uniquely decomposes a general integer $n$ into a multiple $l$ of $M$
and a remainder $\hat{n}\in \mathbb{Z}_M=\{0,1,..,M-1\}$. Accordingly
\ba \label{6.49}
&& Q_M \omega_M^{-1} Q_M\;e^M_{\tilde{n}}
= \sum_{\tilde{n}'}\; e^M_{\tilde{n}'}\;
\sum_{n\not=0}\; \omega(n)^{-1}\; 
\delta_{\tilde{n},\hat{n}}\;\delta_{\tilde{n}',\hat{n}}
\frac{2\;M^2\;[1-\cos(k_M\hat{n})]}{[2\pi n]^2}
\nonumber\\
&=& [1-\delta_{\tilde{n},0}]\;e^M_{\tilde{n}}\;
\sum_{l\in \mathbb{Z}}\; \omega(\tilde{n}+l\; M)^{-1}
\frac{2\;M^2\;[1-\cos(k_M\tilde{n})]}{[2\pi (\tilde{n}+lM)]^2}
\ea
whence for $M> n>0$
\be \label{6.50}
\omega_M(n)^{-1}=
\sum_l\; \omega(n+lM)^{-1}\;\frac{2[1-\cos(k_M n)]}{[k_M (n+lM)]^2}
=\omega(n)^{-1}\; \frac{2[1-\cos(k_M n)]}{[k_M n]^2}
+\sum_{l\not=0}\; \omega(n+lM)^{-1}\;\frac{2[1-\cos(k_M n)]}{[k_M (n+lM)]^2}
\ee
Since $\omega(n)=2\pi |n|$, at fixed $n$ the first term in (\ref{6.50}) 
converges to $\omega^{-1}(n)$ as $M\to \infty$ while the modulus of 
the second is bounded 
by the series 
\be \label{6.51}
\frac{4}{(2\pi)^3\;M}\; \sum_{l=1}^\infty\;[
\frac{1}{[l+\frac{n}{M}]^3} 
+\frac{1}{[l-\frac{n}{M}]^3} 
<    
\frac{4}{(2\pi)^3\;M}\; \sum_{l=1}^\infty\;[
\frac{1}{l^3} 
+\frac{1}{[l-\frac{1}{2}]^3} 
\ee
for $n<M/2$ and thus converges to zero as $M^{-1}$. Accordingly
$\omega_M(n)-\omega(n)=O(1/M)$ at fixed $n$. Then 
(\ref{6.45}) becomes 
\ba \label{6.52}
&& <e_n, [\Delta K]\; e_{n'}>   
=2[\omega(n)\;\delta_{n,n'}-\sum_{\tilde{n}\in \mathbb{Z}_M}\;
\omega_M(\tilde{n})\;\sum_{m_1,m_2}\;
\frac{e^{i k_M n'}-1}{2\pi i n'}\;
[\frac{e^{i k_M n}-1}{2\pi i n}]^\ast\;
e^M_{\tilde{n}-n}(m_1)
e^M_{n'-\tilde{n}}(m_2)
\nonumber\\
&=& 2M^2[\omega(n)\;\delta_{n,n'}-\sum_{\tilde{n}\in \mathbb{Z}_M}\;
\omega_M(\tilde{n})\;
\delta_{\tilde{n},\hat{n}}\delta_{\tilde{n},\hat{n}'}
\frac{2[1-\cos(k_M \tilde{n})}{(2\pi)^2 nn'}
\ea
where $n=\hat{n}+lM,\;n'=\hat{n}'+l'M$ and $\hat{n},\hat{n}'\in \mathbb{Z}_M$.
Since $0>n,n'>-n_0$ and eventually $M>n_0$ we have $l=l'=-1$ and 
$\hat{n}=M+n=\tilde{n},\;\hat{n}'=M+n'=\tilde{n}$ therefore $n=n'$ in the 
second term of (\ref{6.52}) and $\tilde{n}=n+M$
\be \label{6.52}
<e_n, [\Delta K]\; e_{n'}>   
=2\delta_{n,n'}[\omega(n)-\omega_M(M+n)\frac{2[1-\cos(k_M(n))]}{(k_M n)^2}
\ee
Note that for $-M<-n_0<n<0$ we have 
\ba \label{6.53}
&& \omega_M(M+n)^{-1}
=\sum_{l\in \mathbb{Z}}\;\omega(M+n+lM)^{-1}\;
\frac{2[1-\cos(k_M(M+n)]}{[k_M(M+n+l M)]^2}
\nonumber\\
&=& \sum_{l\in \mathbb{Z}}\;\omega(n+lM)^{-1}\;
\frac{2[1-\cos(k_M n)]}{[k_M(n+l M)]^2}
\to \omega(n)^{-1}=\omega(-n)^{-1}
\ea
as $M\to \infty$. Thus indeed (\ref{6.41}) converges to zero. 

Next consider
\be \label{6.54}
\int\;dx\;\int \; dx'\; F(x)\;F(x')\;
[\Delta K]_+(x,x')\;[\Delta K]_+(x,x')
=
\int\;dx\;\int \; dx'\; F(x)\;F(x')\;
[\Delta K]_+(x,x')\;[\Delta K]_-(x',x)
\ee
By the same argument as above, if $F$ has compact momentum support, then 
(\ref{6.54}) is a quadratic polynomial in the $<e_n,[\Delta K]_{n'}>$ 
with $M$ independent coefficients where either $n_0>n,n'>0$ or
$-n_0<n,n'<0$ and hence converges to zero.   

Finally consider  
\be \label{6.56}
\int\; dx\; F(x)\;\int\; dx'\; F(x')\;
g'(x) g'(x')\; [\Delta K]_s(x,x')
\ee
where 
\be \label{6.57}
g'(x)=[Q I_M C_M^{-1} I_M^\dagger C f](x)
\ee
We note that $Q I_M=I_M Q_M,\; [C_M,Q_M]=[C,Q]=0$ implies that 
$I_M^\dagger Q=q-m I_M^\dagger$ whence by the now familar argument  
\be \label{6.57}
g'(x)=[Q I_M C_M^{-1} I_M^\dagger C Q f](x)
\ee
so that 
\ba \label{6.58}
g' &=& \sum_{n,n'\not=0}\sum_{\tilde{n}}\;\omega_M(\tilde{n})\; 
e_{n'}\; <e_{n'}, I_M e^M_{\tilde{n}}>\; 
<I_M e^M_{\tilde{n}}, e_n> \omega(n)^{-1} \hat{f}(n)
\nonumber\\
&=&
\sum_{l,l'\in \mathbb{Z}}\sum_{\tilde{n}}\;\omega_M(\tilde{n})\; 
e_{\tilde{n}+l'M}\; 
\frac{2[1-\cos(k_M \tilde{n}]}{k_M^2(\tilde{n}+lM)(\tilde{n}+l'M)} 
\omega(\tilde{n}+lM)^{-1} \hat{f}(\tilde{n}+l M)
\ea
It follows that $g'$ does not have compact momentum support $n'$ even if 
$f$ does. Therefore $F(x) g'(x)$ also does not have compact momentum support
even if $F$ does. It is not even clear that (\ref{6.58}) converges. This
feature of $f'$ is again due to the fact that the functions $\chi^M_m$ are 
discontinuous. If one would replace them by $\chi^{M,n_0}_m$ where 
$\hat{\chi}^{M,n_0}_m$ is the Fourier expansion of $\chi_m^M$ restricted to 
modes $|n|<n_0$ then $\chi^{M,n_0}_m\to \chi^M_m$ in the $L$ norm 
and if we define $I^{n_0}_M,\; [I^{n_0}_M]^\dagger$ like $I_M,\;I_M^\dagger$
with $\chi^M_m$ replaced by $\chi^{M,n_0}_m$ and first take the limit 
$M\to \infty$ in (\ref{6.56}) and then $n_0\to \infty$ then (\ref{6.56}) 
vanishes as $M\to \infty$. 
This regularisation using the momentum cut-off $n_0$ is similar 
to the zeta function regularisation of section \ref{s3} and 
is justified by 
the following argument: while the $\chi^M_m$ have all the necessary features 
in order to define a renormalisation flow, they are not the only choice. 
There are other, smoother choices \cite{27} satisfying the same necessary 
requirements
but those have a built in compact momentum support of order $M$. In that
case the sum over $l,l'$ in (\ref{6.58}) disappears and the compact 
momentum support of $f$ propagates to that of $g$ and then e.g. $g=Qf$ 
{\it even exacly} for sufficiently large $M$. Then also 
$F g'$ have compact momentum support and the same 
argument 
as was made for (\ref{6.41}) and (\ref{6.54}) can be used to show that 
(\ref{6.56}) converges to zero without any regulator. Since the choice 
of the $\chi^M_m$ is quite arbitrary subject to a minimal set 
of requirements and since one wants to probe functions $f$ of compact 
momentum support using their $I_M I_M^\dagger f$ approximants, 
such a smooth choice of $\chi^M_m$ is simply more 
convenient. With respect to any choice we have convergence of 
$I_M I_M^\dagger\to 1_L$ in the $L_2$ sense but the finite resolution  
approximants have additional smoothness or momentum compactness properties 
while others do not and those additional properties turn out 
to be important in the present convergence analysis. 
The strict proof that with the choice of $I_M$
made in \cite{27} expression (\ref{6.56}) converges to zero is  
given in section 5 of \cite{27} and also provides the argument 
that was missing at the end of the previous section to establish convergence 
of the flow of constraints.\\
\\
We conclude this section with the remark that the functions 
$\chi^M_m$ used in \cite{27} are smooth with compact momentum support and 
that smooth smearing functions $F,f$ of constraints and Weyl elements 
respectively are of rapid decrease in the momentum mode label $n$. Thus with 
respect to those functions all estimates of this section pass through 
without any regularisation and convergence is established.    

\section{Discretised Smearing Functions of the constraints}
\label{s7}

As we have seen, the embeddings $J_M$ do not induce a canonical map
$E_M\; L\to L_M$ such that (we drop the index $s$ for the purpose of this 
section)
\be \label{7.1}
D_M(F):=J_M^\dagger\; D(F)\; J_M=:\tilde{D}_M(E_M F)
\ee
However, we may use the map $E_M:=I_M^\dagger$ to define the family of 
discretised smearing functions $F_M:=I_M^\dagger F$
\be \label{7.2}
\tilde{D}_M(F_M):=J_M^\dagger\; D(I_M F_M)\;J_M=D_M(p_M F)
\ee
where 
\be \label{7.3}
p_M=I_M\; I_M^\dagger:\; L\to L
\ee
is a projection due to isometry $I_M^\dagger I_M=1_{L_M}$. This defines a 
consistent family of quadratic forms in the sense that for any $M<M'$
\be \label{7.4}
J_{MM'}^\dagger\; \tilde{D}_{M'}(I_{MM'} F_M)\; J_{MM'}
=\tilde{D}_M(F_M)
\ee
with $I_{MM'}=I_{M'}^\dagger I_M$ thanks to 
$I_{M'} I_{MM'}=I_M$ and 
$J_{M'} J_{MM'}=J_M$. We can therefore compute 
\be \label{7.5}
[\tilde{D}_M(F_M),\;  \tilde{D}_M(G_M)]  
=J_M^\dagger ([D(p_M F),D(p_M G)]
-D(p_M F)(1-P_M)D(p_M G)    
+D(p_M G)(1-P_M)D(p_M F))J_M
\ee
and modulo the central term we have in our case 
\be \label{7.6}
[D(p_M F),D(p_M G)]=D([p_M F,p_m G])
\ee
The new semaring function in (\ref{7.6}) is given by 
\be \label{7.7}
[p_M F, p_M G]:=[p_M F]'\;[p_M G]-[p_M F] [p_M G]'
=p_M([p_M F, p_M G])+(1-p_M)([p_M F, p_M G])
\ee
Thus (\ref{7.5}) becomes 
\be \label{7.8}
[\tilde{D}_M(F_M), \tilde{D}_M(G_M)]
=\tilde{D}_M(\kappa_M(F_M,G_M))
\ee
modulo the central term and the corrections involving $1_{{\cal H}}-P_M$ and 
$1_L-p_M$. Here the discretised structure functions are 
defined by 
\be \label{7.9}
\kappa_M(F_M,G_M):=I_M^\dagger\; \kappa(I_M F_M, I_M G_M),\;\;
\kappa(F,G)=[F,G]
\ee
which are well defined if the functions $\chi^M_m$ defining $I_M$ are 
sufficiently differentiable. We have already seen in the previous section 
that the correction involving $1_{{\cal H}}-P_M$ converges to zero if 
$F$ has compact momentum support. That is no longer the case for 
$F$ replaced by $p_M F$ if the functions $\chi^M_m$ are step functions
but it is the case when those functions themselves have compact momentum
support as those in \cite{27}. The functions $\chi_M^m$ in general 
span a closed, finite dimensional subspace $V_M\subset L$ and their 
derivatives $[\chi^M_m]'$ may or may not lie in $V_M$ (for the case 
\cite{27} they actually do). However, the products $\chi^M_{m'}\;
[\chi^M_m]'$ are no longer in $V_M$ so that the term proportional 
to $1_L-p_M$ does not vanish automatically. If however $F,G$ have 
compact momentum support then the projections $p_M F$ coincide with 
$F$ for sufficiently large $M$ because $V_M$ roughly involves 
all Fourier modes up to order $|n|\le M$ and thus also $[F,G]$ eventually
lies in $V_M$ and the correction involving $1-p_M$ eventually vanishes.

If $F,G$ do not have compact momentum support but are smooth then their 
Fourier transforms are of rapid decrease in the mode label $n$. 
In this case the terms involving $1_L-p_M$ are not exactly zero for 
sufficiently large $M$ but do converge to zero rapidly. 
Thus we see that with respect to the coarse graining maps of \cite{27}
the correction terms of type $1_{{\cal H}}-P_M, 1_L-p_M$ of the discrete 
Virasoro algebra converge to zero in the weak operator topology of 
$\cal H$ and that in particular the central term of the Virasoro algebra
is correctly reproduced.  

\section{Conclusion and outlook}
\label{s8}

In the present work we have investigated the question whether Hamiltonian 
renormalisation in the sense of \cite{4,6,7,11}, while derived in 
the context of ordinary Hamiltonian systems, can be ``abused'' to 
study also generally covariant Hamiltonian systems with an infinite number
of Hamiltonian constraints rather than a single Hamiltonian. We have 
chosen parametrised field theory on the 1+1 cylinder to test related 
questions where the exact quantum theory is known. 

We have explicitly demonstrated that indeed the general framework of 
\cite{11} can be applied, although the system does not exhibit a common
vacuum vector $\Omega$ for all constraint operators due to the central 
term in the Virasoro algebra. The renormalisation flow indeed finds the 
correct fixed point theory. This enabled us to study the constraint 
algebra at finite resolution. That finite resolution algebra {\it 
generically does not close} (not even when including the central term).
However, it does not close for a simple mathematical reason: The constraints
at finite resolution are forced to map states in the Hilbert space of given
finite resolution to themselves. However, to achieve closure, matrix elements 
with states at higher resolution are neeeded. These are restored as we 
increase the resolution and explains why the failure of closure is 
parametrised by the projection $1_{{\cal H}}-P_M$ where $P_M$ projects 
on the given finite resolution subspace. In that sense the failure to close 
{\it does not represent an anomaly but just a finite size artefact}.
In QFT's which are not exactly solvable one can distinguish between true 
anomalies and these artefacts by studying whether their size decreases as we
increase the resolution.     
 
In addition we could address the question 
if and in what sense smearing functions of constraint operators can or should 
also be discretised when probing them at finite resolution. Namely, while 
it is not necessary or even natural to do so, one can use the coarse graining
map that was employed for reasons of renormalisation also for those smearing
functions. This leads to an additional finite size artefact in the 
finite resolution constraint algebra parametrised by $1_L-p_M$ where now 
$p_M$ projects on smearing functions (rather than Hilbert space states) 
of finite resolution. This is because the commutator of constraints 
is smeared by a bilinear expression in two smearing functions and typically 
derivatives thereof of finite order and those aggregates generically leave 
the subspace $p_M L$. However, again these corrections converge to zero 
as we increase the resolution for coarse graining maps with sufficient
smoothness.

In the convergence proofs that we supplied it was important that the 
functions that define the coarse graining maps of the renormalisation flow 
display sufficient smoothness as otherwise the estimates that were needed 
do not hold: the Fourier transform of a merely piecewise smooth function is 
not of rapid decrease and displays the Gibbs phenomenon at the discontinuities
\cite{30}, i.e. the partial Fourier transform of the function 
at finite resolution has points within the resolution size away from the 
discontinuity which differ from the function by a size {\it independent of}
the resolution.

We will use the lessons learnt for more complicated and physically more 
interesting constrained QFT such as PFT in higher dimensions and the 
$U(1)^3$ model for quantum gravity \cite{31} which present the 
next logical step in the degree of complexity as 
in these models the constraint algebra (hypersurface 
deformation algebra) no longer closes with structure constants but only 
structure functions.

\end{document}